\documentclass[12pt]{article}
\pdfoutput=1

\usepackage[colorlinks = true]{hyperref}

\usepackage{scicite,amssymb}

\usepackage{times}
\usepackage[dvipdfmx]{graphicx}
\usepackage{color}
\usepackage{amsmath}
\usepackage{ulem}
\def\Vec#1{\mbox{\boldmath $#1$}}
\newcommand{\red}[1]{{\color{black} #1}}
\newcommand{\blue}[1]{{\color{black} #1}}

\newcommand{\Mb}{M\"ossbauer }
\newcommand{\ybal}{$\beta$-YbAlB$_4$\ }

\newenvironment{sciabstract}{%
\begin{quote} \bf}
{\end{quote}}

\newcounter{lastnote}

\title{
Observation of a Critical Charge Mode\\ in a Strange Metal 
}

\author
{Hisao Kobayashi,$^{1,2 \ast}$ Yui Sakaguchi,$^{1}$ Hayato Kitagawa,$^{1,2}$ Momoko Oura,$^{1,2}$  \\
 Shugo Ikeda,$^{1,2}$ Kentaro Kuga,$^{3}$ Shintaro Suzuki,$^{3}$ Satoru Nakatsuji,$^{3,4,5,6 \ast}$ \\
Ryo Masuda,$^{2,7}$ Yasuhiro Kobayashi,$^{2,7}$ Makoto Seto,$^{2,7}$ \\
Yoshitaka Yoda,$^{8}$ Kenji Tamasaku,$^{2}$ \\
Yashar Komijani,$^{9,10}$ Premala Chandra,$^{10}$ Piers Coleman$^{10,11 \ast}$\\
\\
\normalsize{$^{1}$Graduate School of Material Science, University of Hyogo, 3-2-1 Koto, Hyogo 678-1297, Japan}\\
\normalsize{$^{2}$RIKEN SPring-8 Center, Hyogo 679-5148, Japan}\\
\normalsize{$^{3}$Institute for Solid State Physics, University of Tokyo, Kashiwa 277-8581, Japan}\\
\normalsize{$^{4}$Department of Physics, University of Tokyo, Hongo, Bunkyo-ku, Tokyo 113-0033, Japan}\\
\normalsize{$^{5}$Trans-scale Quantum Science Institute, University of Tokyo, Bunkyo-ku, Tokyo 113-0033, Japan}\\
\normalsize{$^{6}$Institute for Quantum Matter and Department of Physics and Astronomy,}\\
\normalsize{Johns Hopkins University, Baltimore, Maryland 21218, USA}\\
\normalsize{$^{7}$Institute for Integrated Radiation and Nuclear Science, Kyoto University, Osaka 590-0494, Japan}\\
\normalsize{$^{8}$Japan Synchrotron Radiation Research Institute, Hyogo 679-5198, Japan}\\
\normalsize{$^{9}$Department of Physics, University of Cincinnati, Cincinnati, Ohio 45221-0011, USA}\\ 
\normalsize{$^{10}$Department of Physics and Astronomy, Rutgers University, Piscataway, New Jersey 08854, USA}\\
\normalsize{$^{11}$Hubbard Theory Consortium, Department of Physics,}\\
\normalsize{Royal Holloway, University of London, Egham, Surrey TW20 0EX, UK}\\
\\
\normalsize{$^\ast$To whom correspondence should be addressed; E-mail:  kobayash@sci.u-hyogo.ac.jp, }\\
\normalsize{satoru@phys.s.u-tokyo.ac.jp, coleman@physics.rutgers.edu}
}

\date{\today}

\begin{document} 


\baselineskip24pt


\maketitle 
\newpage


\begin{sciabstract}
Quantum electronic matter has long been understood in terms of two limiting behaviors of electrons: 
one of delocalized metallic states, and the other of localized magnetic states.  
Understanding the strange metallic behavior which develops at the
brink of localization demands new probes of the underlying electronic charge dynamics. 
Using a state-of-the-art technique, synchrotron-radiation-based M$\ddot{\text{o}}$ssbauer spectroscopy, we have studied the longitudinal charge fluctuations 
of the strange metal phase of $\beta$-YbAlB$_4$ as a function of temperature and pressure. 
We find that the usual single absorption peak in the Fermi-liquid regime splits into two peaks upon entering the critical regime.
This spectrum is naturally interpreted as a single nuclear transition, modulated by nearby electronic valence fluctuations 
whose long time-scales are further enhanced, due to the formation of charged polarons. 
Our results represent a direct observation of critical charge fluctuations as a new signature of strange metals. 

\end{sciabstract}

\newpage


The strange metal (SM) is a ubiquitous state of matter found to develop in quantum materials with strong correlations, 
often appearing as a fan-shaped region of the phase diagram centered around an unstable quantum critical (QC) point. 
SMs share many commonalities, most-notably {a logarithmic temperature ($T$) dependence of specific heat $C/T\sim -\log T$, a linear-in-$T$ resistivity {$\rho(T) \sim T$}
 \cite{Stewart01}} and a strong violation of Kohlers law in the magnetotransport \cite{Ong91,Nakatsuji08,Analytis14}.  
These  properties {and their universality} defy the standard concept of quasiparticle excitations {and the conventional wisdom of momentum-relaxation-origin of the conductivity}, central to the Fermi liquid (FL) theory of metals. {This enigma has}
prompted a wide range of possible {origins}, 
including spin-density instability \cite{Stewart01}, Fermi surface instablity \cite{Paschen04,Shishido2005a,QCnphys892}, 
valence quantum criticality \cite{Kuga2018a}, charge stripes \cite{Laliberte}, and nematicity \cite{Fradkin,chu824,lawler2010}  \red{and motivated} novel approaches, including the holographic duality \cite{Holobook2,Holobook1,HartnollMacKenzie21} 
and simulation using cold atoms \cite{Brown19}. 

While the spin dynamics at quantum criticality has been extensively studied, little is known experimentally about the charge
dynamics as appropriate laboratory probes are scarce.
Conventionally, charge dynamics are studied using optical spectroscopy\cite{Prochaska18}, but these methods only probe the \blue{low-momenta,} divergence-free {\it transverse}
components of the current $\Vec{J}= \sigma \Vec{E} \perp \Vec{k}$
that, by the continuity equation, do not couple to fluctuations in the
charge density.
\blue{Longitudinal charge fluctuations can be probed by electron
energy loss spectroscopy (EELS) but are limited to energies above the
Debye energy due to a phonon background\,\cite{Vig2017,Mitrano2018,Husain2019}}. A classic method to detect low frequency longitudinal charge dynamics is M$\ddot{\text{o}}$ssbauer spectroscopy,  
successfully used in the past to detect the slowing of the charge
dynamics at charge ordering transitions of Eu and Fe based compounds
\cite{Berkooz68,Takano79}.

\begin{figure}
 \begin{center}
\vspace*{-1.0cm}
\hspace*{-1.5cm}
\includegraphics [width=0.83\linewidth, clip]  {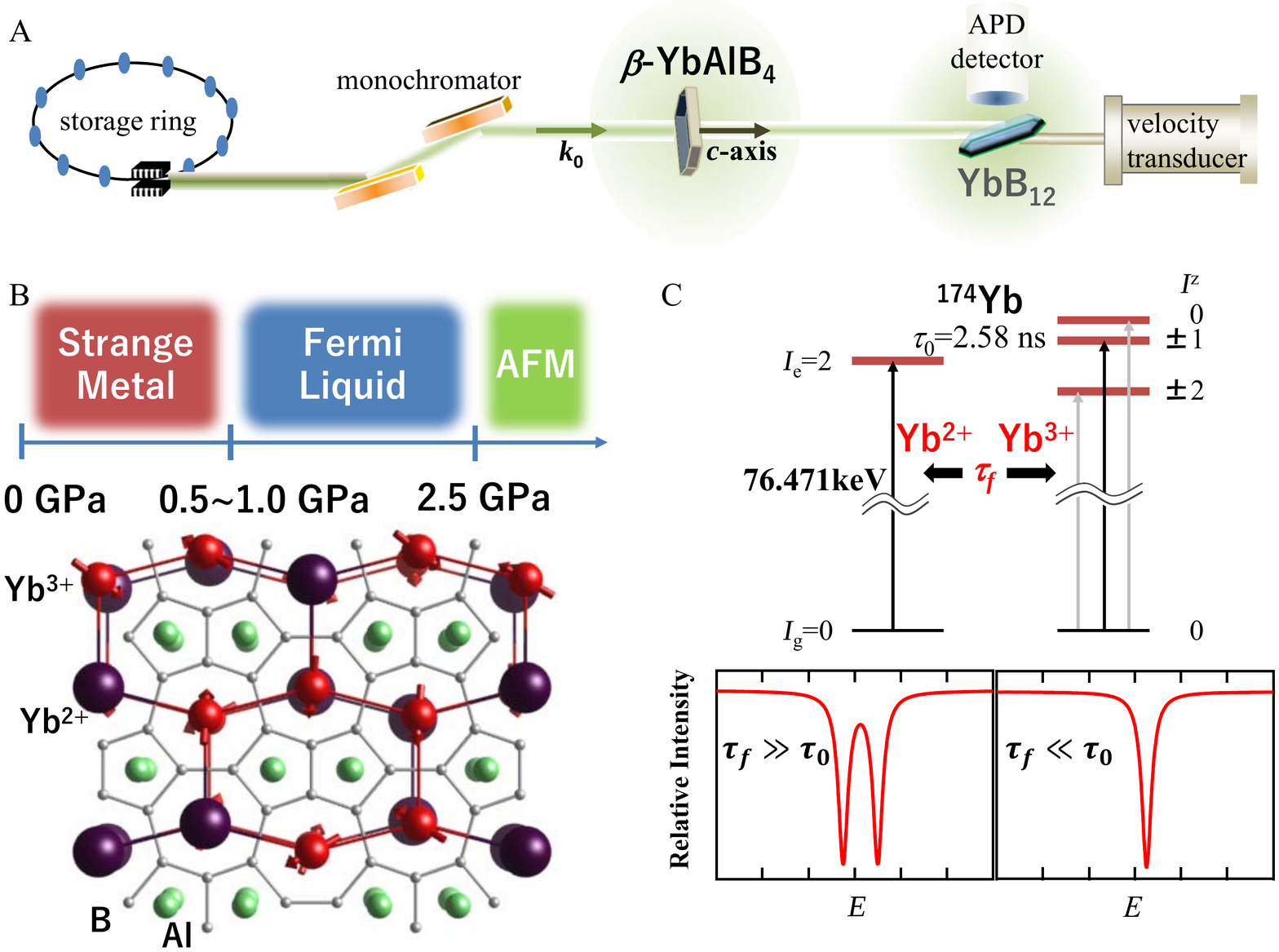}
\vspace*{+.5cm}
   \caption{
(A) Schematic of our experimental setup for the synchrotron-radiation-based $^{174}$Yb M$\ddot{\text{o}}$ssbauer spectroscopy \cite{Masuda14}. 
The $^{174}$Yb nuclear resonance ($E_{\gamma}=$76.471keV) was obtained by synchrotron radiation using a monochromator.
The $c$-axis of the single crystalline $\beta$-YbAlB$_4$ samples was aligned along the propagation vector $\Vec{k}_0$ of the incident X-ray under both ambient and external pressure.
The single-crystalline YbB$_{12}$ were cooled at 26K. 
A Si avalanche photodiode (APD) detector was used to measure delayed incoherent emission from $^{174}$Yb nuclei in the YbB$_{12}$.
(B) Schematic phase diagram of $\beta$-YbAlB$_4$ as a function of pressure at low temperatures (Top) 
and cartoon of the crystal structure of $\beta$-YbAlB$_4$ with a snapshot of the Yb valences, i.e., Yb$^{2+}$ (large blue sphere) and Yb$^{3+}$ (small red sphere with arrow indicating magnetic moment) (Bottom).
(C) (Top) Energy level diagrams of the excited $^{174}$Yb ($I_e$=2) nuclear state with the lifetime of $\tau_0 =2.58$ns surrounded by different charge configurations. 
The allowed M$\ddot{\text{o}}$ssbauer transitions are indicated by arrows, where the black arrows represent two selected transitions for $c \parallel \Vec{k}_0$.
(Bottom) Two typical M$\ddot{\text{o}}$ssbauer absorption spectra at limiting cases with $\tau_f \gg \tau_0$ and $\tau_f \ll \tau_0$ 
where $\tau_f$ is a characteristic timescale of fluctuation between two different charge configurations.
}
\label{Fig:Fig_00}
\end{center}
\end{figure}

However, the widespread adoption of M$\ddot{\text{o}}$ssbauer methods has been long hindered by the lack of suitable radioisotope sources. 
To overcome these difficulties, a new generation of M$\ddot{\text{o}}$ssbauer spectroscopy has recently been developed 
using synchrotron radiation (SR) \cite{Seto09}. 
SR-based M$\ddot{\text{o}}$ssbauer spectroscopy (see Fig.\,\ref{Fig:Fig_00} A) can be used for a wide range of M$\ddot{\text{o}}$ssbauer isotopes,  
providing improved energy resolution for these isotopes with
shorter-lifetimes; it offers an unprecedented capability to select a particular nuclear transition, taking advantage of the perfectly polarized SR. 
This new approach presents an ideal probe to resolve {\it longitudinal} charge dynamics in materials 
for which conventional M$\ddot{\text{o}}$ssbauer techniques are inapplicable.

Here we report the first direct observation of critical charge dynamics in a SM {regime} using SR-based $^{174}$Yb M$\ddot{\text{o}}$ssbauer spectroscopy. 
The heavy fermion metal $\beta$-YbAlB$_4$ provides an ideal platform
to study a SM {regime} at ambient pressure in a stoichiometric
crystal \cite{Nakatsuji08,Matsumoto11}. 
{
In \ybal, core level X-ray studies have  established the presence of
an intermediate
valence state
caused by valence fluctuations  
between two ionic configurations\cite{Okawa10}
Yb$^{2+} \rightleftharpoons$Yb$^{3+}+ e^{-}$. 
Usually, in heavy fermion compounds, such
valence fluctuations are 
too fast to be
observed by \Mb  spectroscopy
\cite{Varma1976,sampathkumaran86,oldmoss,oldmoss2}, but here we show
that this is not the case in the SM.
}

\blue{M\"ossbauer spectroscopy measures the shift in a nuclear absorption line due to changes in the local ($q$-integrated) charge density.  
The characteristic time-scale of the measurement is the lifetime of the nuclear excited state, $\tau_0\sim 2.5$ns in $^{174}$Yb.  
Charge fluctuations that are much shorter in time than $\tau_{0}$ produce a single motionally narrowed absorption line, 
whereas charge fluctuations that are much longer in time than $\tau_{0}$ produce a double peak absorption line, 
corresponding to the two different valence states of the Yb ion (see Figure \ref{Fig:Fig_00} C). 
By fitting the M\"ossbauer absorption line-shape, one can detect charge fluctuations 
with time-scales in the range $\sim$ 0.1$\tau_{0}$  to  $\sim$ 10$\tau_{0}$ \cite{SM}. 
 }

\ybal exhibits quantum criticality without tuning in 
an intermediate
valence state
 \cite{Okawa10}, and the application of 
an infinitesimal magnetic field $B$ tunes the SM into a FL with $k_{\rm B} T_{\rm FL}\sim\mu_{\rm B} B$. 
The {slope of the linear-in-$T$ resistivity} $\rho(T)\sim T$ over $T$ between 0.5 and {25} K at ambient pressure, \red{corresponds to a nearly quantum-saturated scattering rate $\tau^{-1}_{tr}=0.4 \times k_BT/\hbar$ \cite{SM}, thus establishing $\beta$-YbAlB$_4$ as a system with Planckian dissipation \cite{Legros2019}.} \red{This anomalous $\rho(T)$ and its extension} over a broad pressure ($p$) range from ambient pressure to $p^* \sim$ 0.5GPa \cite{Nakatsuji08,Matsumoto11,Tomita15} (see Fig.\,\ref{Fig:Fig_00} B) provides an excellent setting for high precision measurements of the critical charge fluctuations, \red{likely of relevance to the broader family of SMs}.

\begin{figure}
 \begin{center}
\vspace*{-2.5cm}
  \includegraphics [width=1.0\linewidth, angle=0, clip]  {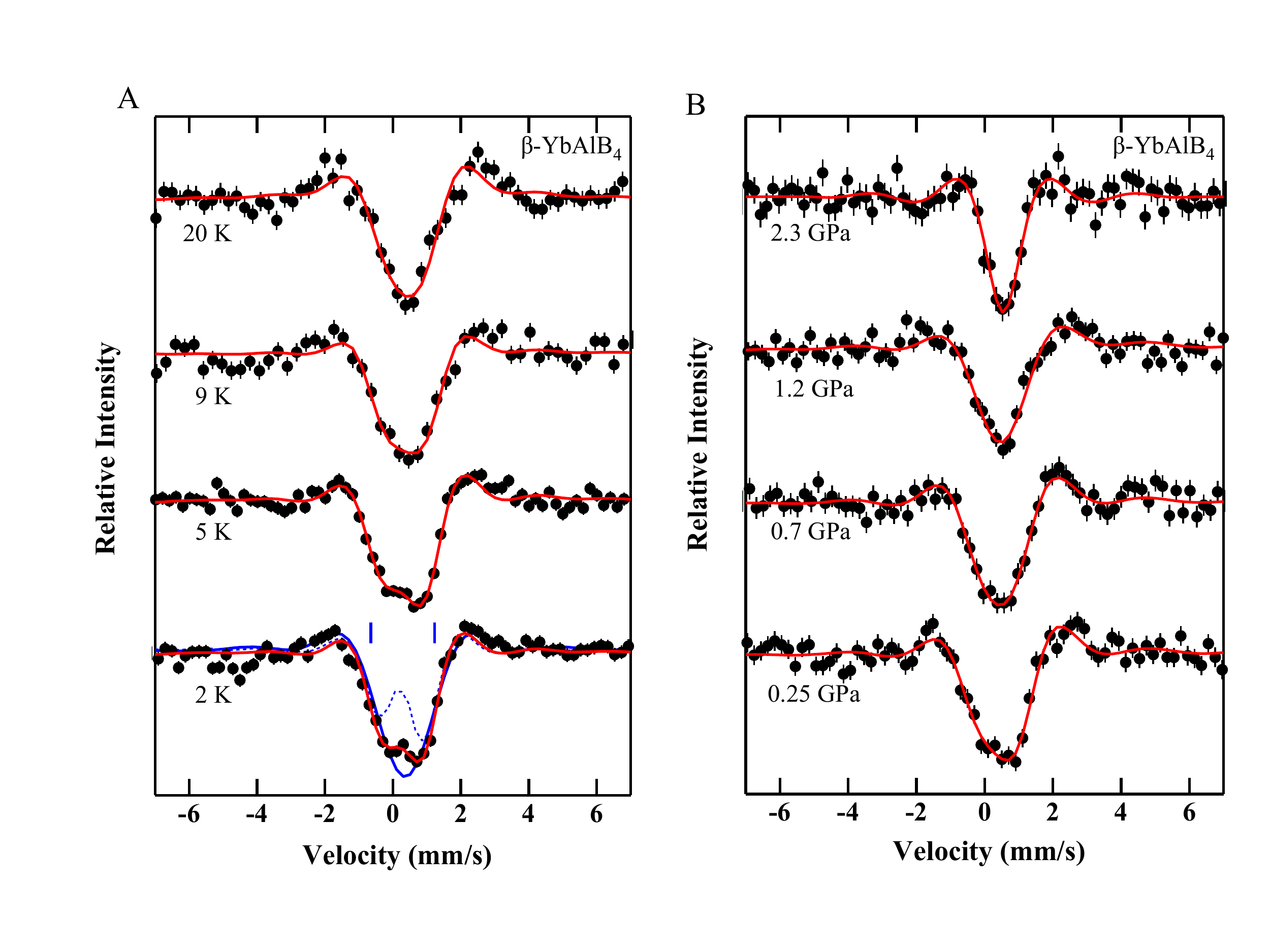}
\vspace*{-0.5cm}
  \includegraphics [width=0.9\linewidth, angle=0, clip]  {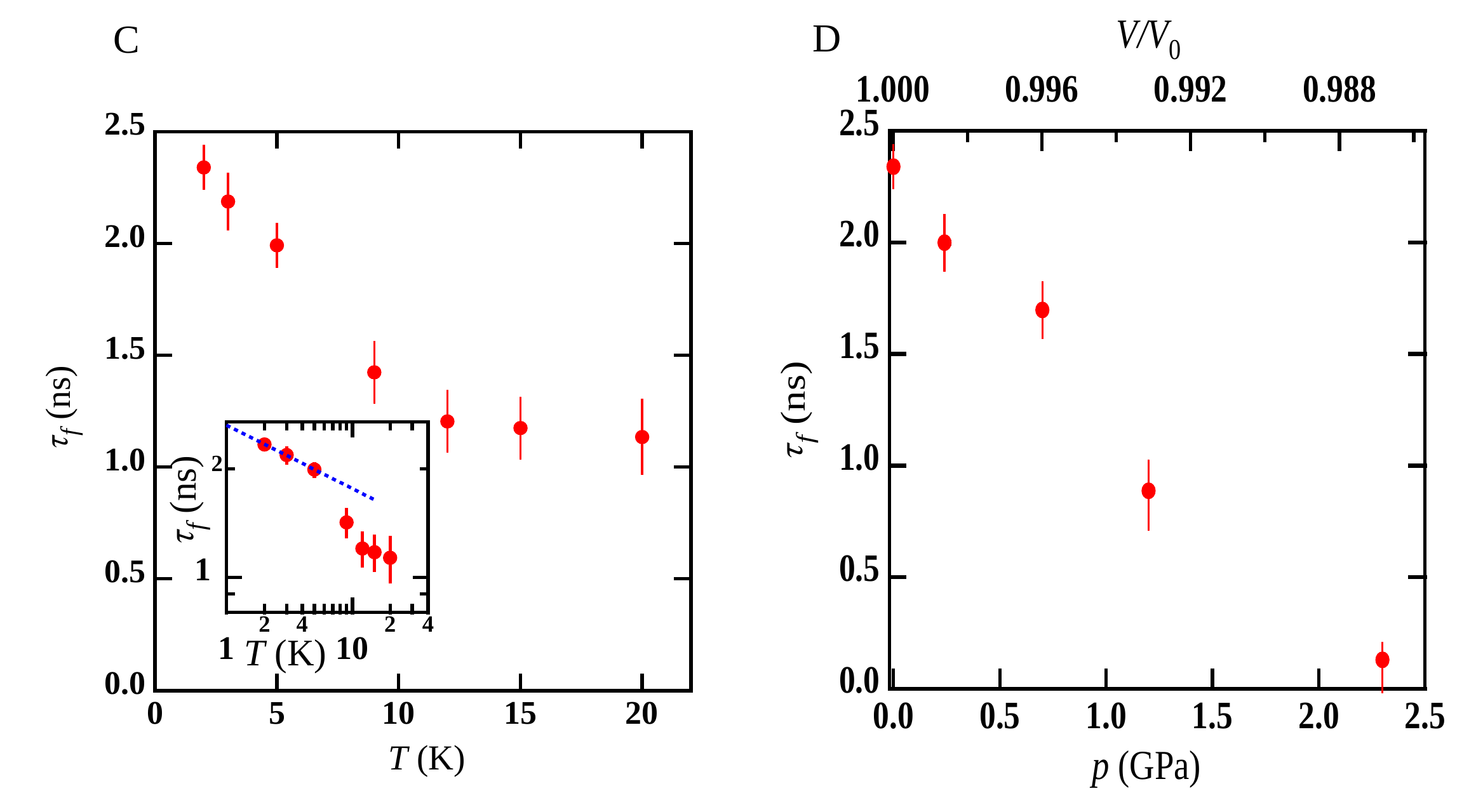}
   \caption{
Selected synchrotron-radiation-based $^{174}$Yb M$\ddot{\text{o}}$ssbauer spectra of $\beta$-YbAlB$_4$ as a function of temperature ($T$) 
at ambient pressure (A) and under external pressure ($p$) at 2K (B).
The $c$-axis of the single crystalline $\beta$-YbAlB$_4$ samples was aligned along the propagation vector $\Vec{k}_0$ of the incident X-ray. 
The solid circles with error bar and the red solid lines present the observed and the analytical spectra, respectively. 
In A, the broken blue line in the spectrum at 2K represents the spectrum with two \red{static} nuclear transitions expected with our experimental energy resolution, \red{whereas the solid blue line shows a fit to the wings of the lineshape, discarding the double-peak structure in the center. The deviation at the center corresponds to $5\sigma$ statistical significance \cite{SM}}.
\red{Temperature} $T$ (C) and \red{pressure} $p$ (D) dependences of the refined fluctuation time $\tau_f$ between two different Yb charge states in $\beta$-YbAlB$_4$. 
 (Inset in (C)) Log-log plots of $\tau_f$ versus $T$ in $\beta$-YbAlB$_4$.
The broken line represents a $\tau_f \sim T^{-0.2}$ relation.
}
\label{Fig:Fig_spec}
\end{center}
\end{figure}

We have investigated how the QC behavior in the SM \red{regime} affects the charge dynamics, following their evolution 
as the SM \red{regime} at ambient pressure transforms into a FL \red{regime} under pressure. 
Above 9K at ambient pressure (Fig.\,\ref{Fig:Fig_spec} A) the M$\ddot{\text{o}}$ssbauer spectra exhibit \blue{a single line} feature.
However, below $T^* \sim 10$K, as one enters the QC region, this  peak broadens into a two-peak structure\red{, with $5\sigma$ significance \cite{SM}}.  
Fig.\,\ref{Fig:Fig_spec} B shows how this two-peak structure observed for $p < $ 0.7GPa at 2K coalesces into a single peak around $p \sim$1.2GPa, 
ultimately sharpening into an almost resolution-limited peak at $p=$ 2.3GPa characteristic of a Fermi liquid \cite{SM}.

The local symmetry at the Yb site of $\beta$-YbAlB$_4$ with the orthorhombic structure allows us to rule out a nuclear origin of the double-peak structure. 
For $c \parallel \Vec{k}_0$ (the propagation vector of the incident X-ray), the symmetry selects two degenerate nuclear transitions $I_g=0 \rightarrow  I_e^z=\pm 1$ from the five $E2$ nuclear transitions ($\Delta I^z  = 0, \pm 1,$ and $\pm 2$) of the $^{174}$Yb M$\ddot{\text{o}}$ssbauer resonance \cite{Hannon88} 
(see Figs.\,\ref{Fig:Fig_00} A and C). 
The absence of magnetic order in $\beta$-YbAlB$_4$ \cite{Matsumoto11,Tomita15} also eliminates magnetic and {non-axially symmetric quadrupolar}
hyperfine  interactions as explanations\cite{SM}. 
This leaves a combination of the electric monopole and axially symmetric quadrupolar
interactions, linking the hyperfine energy to the valence state of the
rare-earth ion, as the only candidate for the observed splitting. 
The presence of a M$\ddot{\text{o}}$ssbauer line splitting then implies a distribution of Yb valences within the crystal. 
We now argue that these result from slow dynamic charge fluctuations.

All Yb sites are crystallographically equivalent in $\beta$-YbAlB$_4$ and SR X-ray diffraction measurements \cite{Sakaguchi16} show 
that the lattice structure does not change up to 3.5GPa at 7K;
\red{furthermore the absence of any low-temperature phase transitions
rules out 
the possibility of a  charge density wave\cite{SM}}. 
Moreover, the residual resistivity ratio (RRR) exceeds 100,  indicating the \blue{low levels of} quenched disorder in this material. \blue{Since disorder broadens the M\"ossbauer absorption peak, our ability to resolve the double-peak structure is consistent with this conclusion.} An attempt to fit the M$\ddot{\text{o}}$ssbauer spectrum with two nuclear transitions (i.e. a static hyperfine interaction), 
using a width corresponding to the experimental energy resolution, fails to reconstruct the feature at 2K and $\sim$ 0 mm/s (blue broken line). 
Thus the two-peak structure and line broadening observed for $T< 5$K and $p<0.7$GPa must derive from a single nuclear transition 
that is dynamically modulated by fluctuations between two different Yb charge states (i.e. a time-dependent hyperfine interactions) (Fig.\,\ref{Fig:Fig_00} C) \cite{SM}.

We have analyzed our M$\ddot{\text{o}}$ssbauer spectra at ambient pressure using a stochastic theory \cite{Anderson54,Kubo54,Blume68_2} 
with a single nuclear transition modulated by two different charge states \cite{SM}.
Fig.\,\ref{Fig:Fig_spec} A shows that the predicted spectra (red lines) well reproduce the two-peak structure in the spectra at low $T$s 
and its subsequent collapse into a single line with increasing $T$.

At ambient pressure, the extracted fluctuation time $\tau_f$ between two different Yb charge states 
is unusually long compared to the electronic timescales, exhibiting a
slow power-law \red{ growth $T^{-\eta }$ ($\eta \sim 0.2$)} on cooling below $T^*$ (Fig.\,\ref{Fig:Fig_spec} C).  
The energy difference between two selected nuclear transitions is almost independent of $T$ up to 20K \cite{SM}, 
so that the development of the two-peak structure in the observed spectra must derive from the marked low-$T$ growth in $\tau_f$. 
On the other hand, as shown in Fig.\,\ref{Fig:Fig_spec} B, the gradual collapse of the two-peak structure in the observed $^{174}$Yb M$\ddot{\text{o}}$ssbauer spectra at 2K with increasing $p$  
indicates that fluctuation timescale $\tau_f$ becomes shorter as a function of $p$. 
The spectra \red{at $p<$1.2GPa} can only be analyzed and reconstructed by the same stochastic model 
used at the ambient pressure, while the spectrum observed at 2.3GPa was simply fit using the static model.
The line-width of this single absorption component was found to be $\Gamma=$ 1.11 mm/s, slightly broader 
than the resolution limit $\Gamma_0=\hbar/\tau_{0}=$ 1.00 mm/s (3mK), for $^{174}$Yb M$\ddot{\text{o}}$ssbauer spectroscopy ($\tau_{0}=$2.58ns). 

As seen in Fig.\,\ref{Fig:Fig_spec} D, $\tau_f$ gradually decreases with increasing $p$, exhibiting a kink across $\sim p^*$ in between 0.5 and 1GPa, 
approaching the resolution limit at 2.3GPa.
This is roughly consistent with previous $\rho (T)$ measurements in $\beta$-YbAlB$_4$ \cite{Tomita15}; 
at \red{$T<$0.5K} and under $p$, $\rho (T)$ displays $\rho \sim T^{\alpha}$ with $\alpha$ = 3/2 below $p^*$  
and \red{further} application of pressure increases the exponent to $\alpha$
= 2, stabilizing a FL state at about 1GPa \cite{Tomita15}.

The above consistency leads us to interpret the split line-shape observed in the M$\ddot{\text{o}}$ssbauer spectra of the SM 
as unusually slow  valence fluctuations between the Yb$^{2+}$ and Yb$^{3+}$ ionic-like states in $\beta$-YbAlB$_4$, 
on a timescale $\tau_f >1$ns that follows \blue{an approximate} power-law growth \blue{$\tau_f\sim T^{-0.2}$} with decreasing temperature below $T^*$. 
The Yb$^{3+}$ ground state is a $J_z=\pm 5/2$ moment as deduced by varying incident angle of the X-ray \cite{SM}.
The slow charge fluctuations extend up to $p^*$, beyond which a conventional valence fluctuation state 
with rapid charge fluctuation takes over in the pressured regime corresponding to the FL \red{regime}.

\begin{figure}
 \begin{center}
  \includegraphics [width=1.0\linewidth, angle=0, clip]  {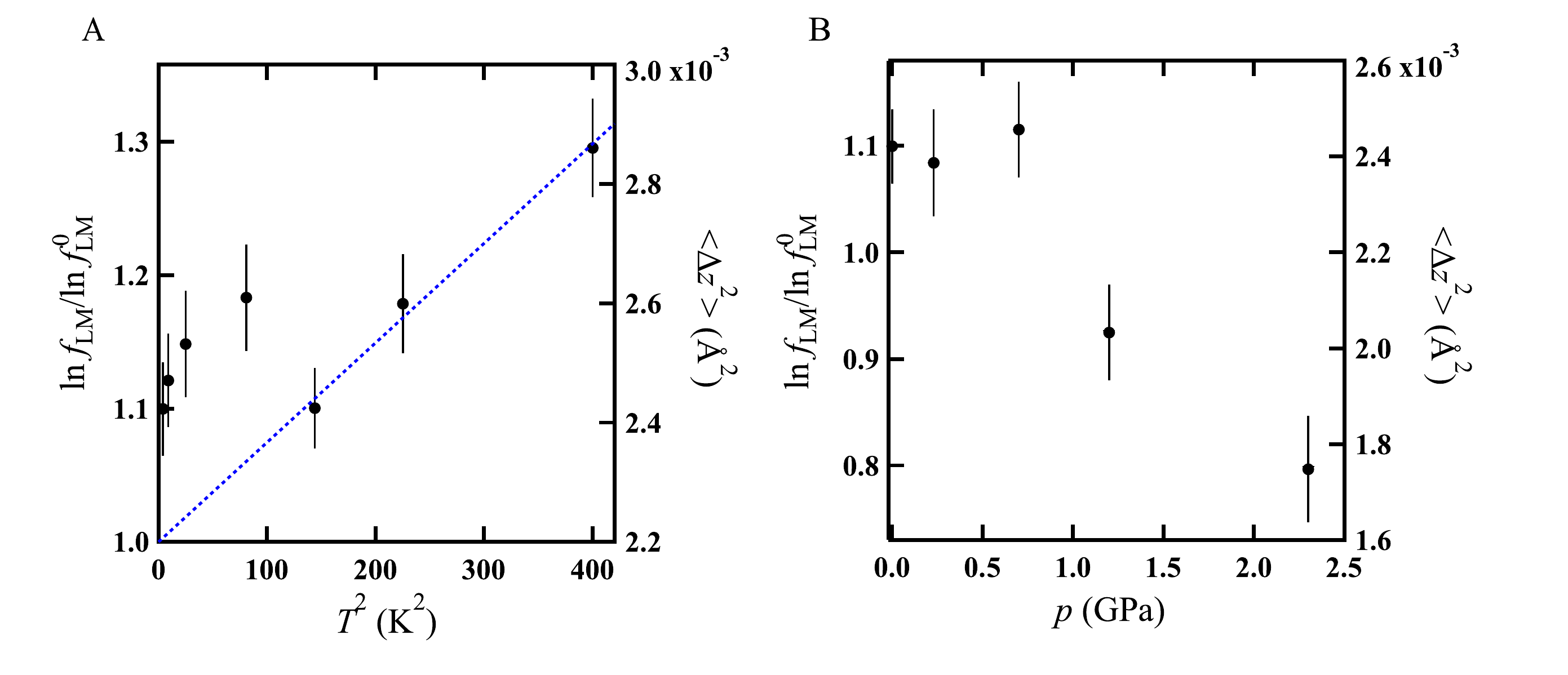}
   \caption{
Lamb-M$\ddot{\text{o}}$ssbauer factor $f_{\rm LM}$ ($\ln f_{\rm LM} /\ln f_{\rm LM}^{0} $) as a function of $T^2$ at ambient pressure (A) and under external pressure $p$ at 2K (B) for $\beta$-YbAlB$_4$.
In (A), the broken line represents a linear relation between $\ln f_{\rm LM}$ and $T^2$.
In (A) and (B), $\ln f_{\rm LM}^{0}  (\propto - \frac{3}{2} \frac{E_{\rm R}}{k_{\rm B} \Theta_{\rm D}^{\rm{Yb}} })$ was estimated above 12K at ambient pressure.
For $^{174}$Yb M\"ossbauer resonance of  $k_0$=38.75\AA$^{-1}$, $\langle \Delta z^2 \rangle$ for the Yb ions was evaluated in $\beta$-YbAlB$_4$ 
from the $T$ and $p$ dependences of $\ln f_{\rm LM}/\ln f_{\rm LM}^{0}$ using $\Theta_{\rm D}^{\rm{Yb}}$= 95K.
In (A) and (B), the $\langle \Delta z^2 \rangle$ values (right axis) are $\sim 2.6 \times 10^{-3}$\AA$^2\,$ 
in the SM \red{regime} and decrease to $1.7 \times 10^{-3}$\AA$^2\,$ in the pressured regime corresponding to the FL \red{regime}, 
which is comparable with that for YbAl$_2$ \cite{Weschenfelder83}.
}
\label{Fig:Fig_LM}
\end{center}
\end{figure}

The unusual aspect of the observed charge dynamics is that not only are they slower than the Planckian time \red{$\tau_f\gg \tau_{tr}\sim 10^{-2}$\,ns at 2K}, 
but they are also slower than the characteristic time-scale of the lattice vibrations as we will show shortly. 
In this situation, the lattice is expected to adiabatically respond to the associated charge redistribution. Each valence fluctuation of Yb atoms is then dressed 
by $N_p$ phonons, leading to the formation of a polaron \cite{Sherington76,Hewson1979} and renormalizing the matrix element for the charge fluctuations and providing a mechanism for enhancing their time-scale  ($\tau_f\to \tau_fe^{N_p}$) \cite{SM}. 
Analysis of the M$\ddot{\text{o}}$ssbauer spectra allows us to directly check this scenario.
We have used the $T$-dependence of the absorption components in the spectra 
to determine the Lamb-M$\ddot{\text{o}}$ssbauer (recoil-free) factor $f_{\rm LM}$ in $\beta$-YbAlB$_4$, 
the equivalent of the Debye-Waller factor in a usual scattering experiment. 
Generally, $-\ln(f_{\rm LM}) = k_0^2 \langle \Delta z^2 \rangle$, where $\Delta z$ is an atomic displacement from a regular position 
in a crystal along the direction of $\Vec{k}_0$ \cite{Trammell62}.
\blue{ 
The expression for the variance in atomic position is
\begin{equation}\label{Deltaz}
\langle \Delta z^{2}\rangle 
\propto
 \int_{0}^{\infty  } d\omega \frac{F(\omega)}{\omega}
\overbrace {
\left[\frac{1}{2}+ \frac{1}{e^{\omega/T}-1} \right]}^{{\frac{1}{2}\coth (\beta \omega /2)}}
\end{equation}
where $F(\omega)$ is the (partial) phonon density of states.} 
\blue{In a Debye model, $F(\omega)\propto \omega^{2}$ which leads to $\langle \Delta z^{2}\rangle \propto \bigl [ 3/2 + (\pi/\Theta_{D})^{2}T^{2 }\bigr ]$ at $T \ll \Theta_{\rm D}$, where $\Theta_{\rm D}$ is the Debye temperature \cite{SM}. 
As seen in Fig.\,\ref{Fig:Fig_LM} A, this Deybe relation
holds above $T^{*}$ at ambient pressure, 
where $\tau_f$ ($\sim$ 1.15ns) is independent of $T$; from this we estimate $\Theta_{\rm D}$ = 95K, 
corresponding to the lattice response time $\tau_L \sim {h}/k_{\rm B} \Theta_{\rm D} \sim 0.5$ps, 
so that $\tau_f \gg \tau_L$. The estimated $\Theta_{\rm D}$ (= 95K) value is unusually smaller than 
that (195K) of a conventional valence fluctuation metal YbAl$_2$ \cite{Weschenfelder83}.
This indicates that the lattice vibrations are soft in $\beta$-YbAlB$_4$, suggesting an enhanced effective coupling
between slow charge fluctuation modes and lattice vibrations. 

Additionally, we see from Fig.\,\ref{Fig:Fig_LM}A that in the QC regime below $T^*$, where $\tau_f$ develops temperature-dependence, $\langle \Delta z^2\rangle$ departs from this Debye behavior, indicating an enhancement in the quantum fluctuations, $\langle \Delta z^2 \rangle = \langle \Delta z^2 \rangle_{\rm Debye} + \delta\langle \Delta z^2 \rangle$, of the Yb ions. 
Notably, the $\sqrt{\delta \langle\Delta z^2\rangle} \sim 0.014$\AA~  rms fluctuation observed here is comparable to the quantum fluctuations of the phonon mode, 
around $0.05$\AA~ estimated from $\frac{1}{4} k_{\rm B}
\Theta_{\rm D} \sim \frac{1}{2} m_{\rm Yb} (k_{\rm B} \Theta_{\rm
D}/\hbar)^2 \langle{\Delta z^2}\rangle$. 
Fig.\,\ref{Fig:Fig_LM} B shows that $\langle \Delta z^2\rangle$ is approximately constant at 2K for $p < p^*$ and then drops when $p>p^*$, 
indicating that the anomalous vibrations of the lattice, $\delta \langle \Delta z^2 \rangle$, disappear in the FL {regime} at low temperatures.

The saturation of 
$\langle \Delta z^{2}\rangle $ for  $T<T^{*}$ and $p<p^{*}$
implies that 
the phonon spectrum $F (\omega)$ has changed its form to
compensate the $\coth (\beta \omega/2)$ term in the
integral \eqref{Deltaz}. This then suggests that at energies and temperatures below
$T^{*}$, $F (\omega )$ acquires a temperature-dependence $F (\omega, T)= \phi(\omega)  \tanh (\omega/2T)$ that cancels the $\coth(\omega/2T)$ term in integral \eqref{Deltaz}.
The function $\tanh(\omega/2T)\sim\omega/2T$ for $\omega\ll T$ and
$\tanh(\omega/2T)\sim 1$ for $\omega\gg T$, and thus has the Marginal
Fermi Liquid (MFL) form. This enhancement in phonon density of states should be observable in inelastic neutron scattering} \blue{measurements.
\red{Since the phonons are linearly coupled to the charge density of the electrons, the appearance of 
a MFL component in the phonon spectrum is an indication of MFL behavior in the charge fluctuations.}
}
The enhancement of $\tau_f$ by polaron formation has been crucial for slowing the charge fluctuations down to time-scales accessible to M$\ddot{\text{o}}$ssbauer spectroscopy.

A possible interpretation of our results is the quantum critical tuning of a critical end point of a classical valence transition \cite{Watanabe10} 
between the Yb$^{2+}$  and Yb$^{3+}$ ionic states. 
Such first order valence transition lines, with second order end-points, are well established in rare earth compounds. 
It has been suggested\cite{Watanabe10} that the tuning of such an endpoint to zero temperature may provide an explanation of the observed M$\ddot{\text{o}}$ssbauer spectra. 

An alternative interpretation is that the observed valence fluctuation modes are an intrinsic property of the SM {regime} 
connected with a spin charge separation that develops with the collapse of the $f$-electron Fermi surface \cite{Oshikawa00,Senthil03,Pixley2012,Komijani19}.
This scenario suggests that similar slow charge fluctuations will be manifested in the M$\ddot{\text{o}}$ssbauer spectra of any partial Mott localization critical point, 
e.g. in other heavy-fermions and Iron-based superconductors. 

In summary we have provided direct evidence for unusually slow charge fluctuations in the SM \red{regime} of $\beta$-YbAlB$_4$ 
using a state-of-the-art technique, SR-based M$\ddot{\text{o}}$ssbauer spectroscopy. 
Because their time-scales are longer than that of the lattice
response, we have \red{inferred} polaronic formation in the mixed
valence regime\cite{Sherington76,Hewson1979}. 
Both the slow charge fluctuation modes and the anomalous vibrations of the lattice disappear in the pressure-induced FL regime. 
An interesting possibility is that these observed slow charge modes are the origin of the linear resistivity often observed in SMs. 
Various theoretical approaches \cite{Holobook2,Holobook1} have suggested that the novel transport properties of SMs are linked 
to the universal quantum hydrodynamics of a Planckian
metal. \red{Since the local equilibrium is established at the scale of
Planckian time, it is natural to regard} the slow charge fluctuations
detected here as a \red{possible} signature of \red{{a new hydrodynamic mode}}. 
This would lead us to expect that nano-second charge fluctuations and anomalous vibrations are not unique to $\beta$-YbAlB$_4$, 
but rather, are universal properties of SM \red{regime}s in quantum materials.

\newpage

\clearpage

\clearpage

\clearpage

\clearpage
\bibliography{YbAlB4Ref}

\begin{thebibliography}{10}

\bibitem{Stewart01}
G.~R. Stewart, {\it Rev. Mod. Phys.\/} {\bf 73}, 797 (2001).

\bibitem{Ong91}
T.~R. Chien, Z.~Z. Wang, N.~P. Ong, {\it Phys. Rev. Lett.\/} {\bf 67}, 2088
  (1991).

\bibitem{Nakatsuji08}
S.~Nakatsuji, {\it et~al.\/}, {\it Nature Physics\/} {\bf 4}, 603 (2008).

\bibitem{Analytis14}
J.~G. Analytis, {\it et~al.\/}, {\it Nature Physics\/} {\bf 10}, 194 (2014).

\bibitem{Paschen04}
S.~Paschen, {\it et~al.\/}, {\it Nature\/} {\bf 432}, 881 (2004).

\bibitem{Shishido2005a}
H.~Shishido, R.~Settai, H.~Harima, Y.~{\=O}nuki, {\it J. Phys. Soc. Jpn.\/}
  {\bf 74}, 1103 (2005).

\bibitem{QCnphys892}
P.~Gegenwart, Q.~Si, F.~Steglich, {\it Nature Phys.\/} {\bf 4}, 186 (2008).

\bibitem{Kuga2018a}
K.~Kuga, {\it et~al.\/}, {\it Sci. Adv.\/} {\bf 4}, eaao3547 (2018).

\bibitem{Laliberte}
F.~Lalibert\'{e}, {\it et~al.\/}, {\it Nature Communications\/} {\bf 2}, 432
  (2011).

\bibitem{Fradkin}
E.~Fradkin, S.~A. Kivelson, M.~J. Lawler, J.~P. Eisenstein, A.~P. Mackenzie,
  {\it Annual Review of Condensed Matter Physics\/} {\bf 1}, 153 (2010).

\bibitem{chu824}
J.-H. Chu, {\it et~al.\/}, {\it Science\/} {\bf 329}, 824 (2010).

\bibitem{lawler2010}
M.~Lawler, {\it et~al.\/}, {\it Nature\/} {\bf 466}, 347,351 (2010).

\bibitem{Holobook2}
J.~Zaanen, Y.-W. Sun, Y.~Liu, K.~Schalm, {\it Holographic Duality in Condesed
  Matter Physics\/} (Cambridge University Press, 2016).

\bibitem{Holobook1}
S.~A. Hartnoll, A.~Lucas, S.~Sachdev, {\it Holographic Quantum Matter\/} (MIT
  Press, 2018).

\bibitem{HartnollMacKenzie21}
S.~A. Hartnoll, A.~P. Mackenzie, {\it ArXiv:2107.07802\/}  (2021).

\bibitem{Brown19}
P.~T. Brown, {\it et~al.\/}, {\it Science\/} {\bf 363}, 379 (2019).

\bibitem{Prochaska18}
L.~Prochaska, {\it et~al.\/}, {\it arxiv: 1808.02296 (2018).\/}  (2018).

\bibitem{Vig2017}
S.~Vig, {\it et~al.\/}, {\it SciPost\/} {\bf 3}, 026 (2017).

\bibitem{Mitrano2018}
M.~Mitrano, {\it et~al.\/}, {\it Proc Natl Acad Sci USA\/} {\bf 115}, 5392
  (2018).

\bibitem{Husain2019}
A.~A. Husain, {\it et~al.\/}, {\it Physical Review X\/} {\bf 9}, 041062 (2019).

\bibitem{Berkooz68}
O.~Berkooz, M.~Malamud, S.~Shtrikman, {\it Solid State Communications\/} {\bf
  6}, 185 (1968).

\bibitem{Takano79}
M.~Takano, N.~Nakanishi, Y.~Takeda, S.~Naka, {\it J. Phys. Colloques\/} {\bf
  40}, C2 (1979).

\bibitem{Seto09}
M.~Seto, {\it et~al.\/}, {\it Phys. Rev. Lett.\/} {\bf 102}, 217602 (2009).

\bibitem{Matsumoto11}
Y.~Matsumoto, {\it et~al.\/}, {\it Science\/} {\bf 331}, 316 (2011).

\bibitem{Okawa10}
M.~Okawa, {\it et~al.\/}, {\it Phys. Rev. Lett.\/} {\bf 104}, 247201 (2010).

\bibitem{Varma1976}
C.~M. Varma, {\it Rev. Mod. Phys.\/} {\bf 48}, 219 (1976).

\bibitem{sampathkumaran86}
E.~V. Sampathkumaran, {\it Hyperfine Interactions\/} {\bf 27}, 183 (1986).

\bibitem{oldmoss}
R.~L. Cohen, M.~Eibsch\"utz, K.~W. West, {\it Phys. Rev. Lett.\/} {\bf 24}, 383
  (1970).

\bibitem{oldmoss2}
I.~Nowik, {\it Hyperfine interactions\/} {\bf 13}, 89 (1983).

\bibitem{SM}
{\it {S}upplementary material is available on {\it Science} online\/} .

\bibitem{Legros2019}
A.~Legros, {\it et~al.\/}, {\it Nature Physics\/} {\bf 15}, 142 (2019).

\bibitem{Tomita15}
T.~Tomita, K.~Kuga, Y.~Uwatoko, P.~Coleman, S.~Nakatsuji, {\it Science\/} {\bf
  349}, 506 (2015).

\bibitem{Hannon88}
J.~P. Hannon, G.~T. Trammell, M.~Blume, D.~Gibbs, {\it Phys. Rev. Lett.\/} {\bf
  61}, 1245 (1988).

\bibitem{Sakaguchi16}
Y.~Sakaguchi, {\it et~al.\/}, {\it J. Phys. Soc. Jpn.\/} {\bf 85}, 023602
  (2016).

\bibitem{Anderson54}
P.~W. Anderson, {\it J. Phys. Soc. Japan\/} {\bf 9}, 316 (1954).

\bibitem{Kubo54}
R.~Kubo, {\it J. Phys. Soc. Japan\/} {\bf 9}, 935 (1954).

\bibitem{Blume68_2}
M.~Blume, {\it Phys. Rev.\/} {\bf 174}, 351 (1968).

\bibitem{Sherington76}
D.~Sherrington, P.~Riseborough, {\it J. Phys. Colloques\/} {\bf 37}, C4 (1976).

\bibitem{Hewson1979}
A.~C. Hewson, D.~M. Newns, {\it Journal of Physics C: Solid State Physics\/}
  {\bf 12}, 1665 (1979).

\bibitem{Trammell62}
G.~T. Trammell, {\it Phys. Rev.\/} {\bf 126}, 1045 (1962).

\bibitem{Weschenfelder83}
D.~Weschenfelder, {\it et~al.\/}, {\it Hyperfine Interactions\/} {\bf 16}, 743
  (1983).

\bibitem{Watanabe10}
S.~Watanabe, K.~Miyake, {\it Phys. Rev. Lett.\/} {\bf 105}, 186403 (2010).

\bibitem{Oshikawa00}
M.~Oshikawa, {\it Phys. Rev. Lett.\/} {\bf 84}, 3370 (2000).

\bibitem{Senthil03}
T.~Senthil, S.~Sachdev, M.~Vojta, {\it Phys. Rev. Lett.\/} {\bf 90}, 216403
  (2003).

\bibitem{Pixley2012}
J.~H. Pixley, S.~Kirchner, K.~Ingersent, Q.~Si, {\it Phys. Rev. Lett.\/} {\bf
  109}, 086403 (2012).

\bibitem{Komijani19}
Y.~Komijani, P.~Coleman, {\it Phys. Rev. Lett.\/} {\bf 122}, 217001 (2019).

\bibitem{Masuda14}
R.~Masuda, {\it et~al.\/}, {\it Appl. Phys. Lett.\/} {\bf 104}, 082411 (2014).

\end{thebibliography}


\begin{thebibliography}{10}

\bibitem{Anderson54}
P.~W. Anderson, {\it J. Phys. Soc. Japan\/} {\bf 9}, 316 (1954).

\bibitem{Kubo54}
R.~Kubo, {\it J. Phys. Soc. Japan\/} {\bf 9}, 935 (1954).

\bibitem{Blume68_1}
M.~Blume, J.~A. Tjon, {\it Phys. Rev.\/} {\bf 165}, 446 (1968).

\bibitem{Blume68_2}
M.~Blume, {\it Phys. Rev.\/} {\bf 174}, 351 (1968).

\bibitem{Komijani16}
Y.~Komijani, P.~Coleman, {\it Phys. Rev. B\/} {\bf 94}, 085113 (2016).

\bibitem{Varma1976}
C.~M. Varma, {\it Rev. Mod. Phys.\/} {\bf 48}, 219 (1976).

\bibitem{sampathkumaran86}
E.~V. Sampathkumaran, {\it Hyperfine Interactions\/} {\bf 27}, 183 (1986).

\bibitem{Okawa10}
M.~Okawa, {\it et~al.\/}, {\it Phys. Rev. Lett.\/} {\bf 104}, 247201 (2010).

\bibitem{oldmoss}
R.~L. Cohen, M.~Eibsch\"utz, K.~W. West, {\it Phys. Rev. Lett.\/} {\bf 24}, 383
  (1970).

\bibitem{oldmoss2}
I.~Nowik, {\it Hyperfine interactions\/} {\bf 13}, 89 (1983).

\bibitem{Ruby74}
L.~Ruby, S., {\it J. Phys. Colloques\/} {\bf 35}, C6 (1974).

\bibitem{Seto09}
M.~Seto, {\it et~al.\/}, {\it Phys. Rev. Lett.\/} {\bf 102}, 217602 (2009).

\bibitem{Masuda14}
R.~Masuda, {\it et~al.\/}, {\it Appl. Phys. Lett.\/} {\bf 104}, 082411 (2014).

\bibitem{Smirnov07}
G.~V. Smirnov, {\it et~al.\/}, {\it Phys. Rev. A\/} {\bf 76}, 043811 (2007).

\bibitem{Lynch60}
F.~J. Lynch, R.~E. Holland, M.~Hamermesh, {\it Phys. Rev.\/} {\bf 120}, 513
  (1960).

\bibitem{Hamillt68}
D.~W. Hamill, G.~R. Hoy, {\it Phys. Rev. Lett.\/} {\bf 21}, 724 (1968).

\bibitem{Seto10}
M.~Seto, {\it et~al.\/}, {\it Journal of Physics: Conference Series\/} {\bf
  217}, 012002 (2010).

\bibitem{Macaluso07}
R.~T. Macaluso, {\it et~al.\/}, {\it Chem. Mater.\/} {\bf 19}, 1918 (2007).

\bibitem{Hannon88}
J.~P. Hannon, G.~T. Trammell, M.~Blume, D.~Gibbs, {\it Phys. Rev. Lett.\/} {\bf
  61}, 1245 (1988).

\bibitem{Matsuda13}
Y.~H. Matsuda, {\it et~al.\/}, {\it Journal of the Korean Physical Society\/}
  {\bf 62}, 1778 (2013).

\bibitem{Ofer68}
S.~Ofer, I.~Nowik, S.~G. Cohen, {\it Chemical Applications of
  M$\ddot{\text{o}}$ssbauer Spectroscopy\/} (Academic Press, New York, p. 427,
  1968).

\bibitem{Freeman62}
A.~J. Freeman, R.~E. Watson, {\it Phys. Rev.\/} {\bf 127}, 2058 (1962).

\bibitem{Meyer79}
{Meyer, C.}, {Gros, Y.}, {Hartmann-Boutron, F.}, {Capponi, J.J.}, {\it J. Phys.
  France\/} {\bf 40}, 403 (1979).

\bibitem{Sakaguchi16}
Y.~Sakaguchi, {\it et~al.\/}, {\it J. Phys. Soc. Jpn.\/} {\bf 85}, 023602
  (2016).

\bibitem{Wolfgang}
W.~Sturhahn, {\it Journal of Physics: Condensed Matter\/} {\bf 16}, S497
  (2004).

\bibitem{Matsumoto11}
Y.~Matsumoto, {\it et~al.\/}, {\it Science\/} {\bf 331}, 316 (2011).

\bibitem{Tomita15}
T.~Tomita, K.~Kuga, Y.~Uwatoko, P.~Coleman, S.~Nakatsuji, {\it Science\/} {\bf
  349}, 506 (2015).

\bibitem{UPd3}
H.~C. Walker, {\it et~al.\/}, {\it Phys. Rev. Lett.\/} {\bf 97}, 137203 (2006).

\bibitem{ceb6}
K.~Hanzawa, T.~Kasuya, {\it Journal of the Physical Society of Japan\/} {\bf
  53}, 1809 (1984).

\bibitem{Winkelmann98}
H.~Winkelmann, {\it et~al.\/}, {\it Phys. Rev. Lett.\/} {\bf 81}, 4947 (1998).

\bibitem{Henning71}
W.~Henning, G.~Bahre, P.~Kienle, {\it Zeitschrift fur Physik A Hadrons and
  nuclei\/} {\bf 241}, 138 (1971).

\bibitem{Ogura06}
M.~Ogura, H.~Akai, {\it Journal of Physics: Condensed Matter\/} {\bf 17}, 5741
  (2005).

\bibitem{Akai17}
H.~Akai. Private communication.

\bibitem{Nevidomskyy09}
A.~H. Nevidomskyy, P.~Coleman, {\it Phys. Rev. Lett.\/} {\bf 102}, 077202
  (2009).

\bibitem{Matsumoto15}
Y.~Matsumoto, {\it et~al.\/}, {\it J. Phys. Soc. Jpn.\/} {\bf 84}, 024710
  (2015).

\bibitem{Ramires12}
A.~Ramires, P.~Coleman, A.~H. Nevidomskyy, A.~M. Tsvelik, {\it Phys. Rev.
  Lett.\/} {\bf 109}, 176404 (2012).

\bibitem{Kuga19}
K.~Kuga, {\it et~al.\/}, {\it Phys. Rev. Lett.\/} {\bf 123}, 036404 (2019).

\bibitem{Nakatsuji08}
S.~Nakatsuji, {\it et~al.\/}, {\it Nature Physics\/} {\bf 4}, 603 (2008).

\bibitem{Legros2019}
A.~Legros, {\it et~al.\/}, {\it Nature Physics\/} {\bf 15}, 142 (2019).

\bibitem{OFarrell2012}
E.~C.~T. O'Farrell, Y.~Matsumoto, S.~Nakatsuji, {\it Physical Review Letters\/}
  {\bf 109}, 176405 (2012).

\bibitem{Pepin07}
C.~P\'epin, {\it Phys. Rev. Lett.\/} {\bf 98}, 206401 (2007).

\bibitem{Pixley2012}
J.~H. Pixley, S.~Kirchner, K.~Ingersent, Q.~Si, {\it Phys. Rev. Lett.\/} {\bf
  109}, 086403 (2012).

\bibitem{Komijani19}
Y.~Komijani, P.~Coleman, {\it Phys. Rev. Lett.\/} {\bf 122}, 217001 (2019).

\bibitem{langfirsov}
I.~G. Lang, Y.~A. Firsov, {\it JETP\/} {\bf 16}, 1301 (1962).

\bibitem{Sherington76}
D.~Sherrington, P.~Riseborough, {\it J. Phys. Colloques\/} {\bf 37}, C4 (1976).

\bibitem{Hewson1979}
A.~C. Hewson, D.~M. Newns, {\it Journal of Physics C: Solid State Physics\/}
  {\bf 12}, 1665 (1979).

\end{thebibliography}
\bibliographystyle{science}

We would like to thank M. Takigawa for very useful discussions and F. Iga for preparation of single-crystalline YbB$_{12}$.
The SR-based $^{174}$Yb M$\ddot{\text{o}}$ssbauer experiments were performed at BL09XU and BL19LXU on SPring-8 
with the approval of the Japan Synchrotron Radiation Research Institute (JASRI) (Proposal Nos. 2011A1450, 2012B1521, 2013B1393, 2015A1458, 2016A1363, and 2019B1597) 
and RIKEN (Proposal Nos. 2016110, 20170019, 20180019, and 20190025). 
This work is partially supported by Grants-in-Aids for Scientific Research on Innovative Areas (15H05882 and 15H05883) from the Ministry of Education, Culture, Sports, Science, and Technology of Japan,  by CREST (JPMJCR18T3), Japan Science and Technology Agency, and by Grants-in-Aid for Scientific Research (15K05182, 16H02209, 16H06345, 19H00650, and 23102723) from the Japanese Society for the Promotion of Science (JSPS), by the Canadian Institute for Advanced Research, the
National Science Foundation grant DMR-1830707 (P. Coleman and
Y. K) and by the U. S. Department of Energy (DOE), Office of Science,
Basic Energy Sciences under award DE-SC0020353 (P. Chandra). 
The Institute for Quantum Matter, an Energy Frontier Research Center was funded by DOE, Office of Science, Basic Energy Sciences under Award \# DE-SC0019331.
P. Chandra and P. Coleman thank S. Nakatsuji and the Institute for Solid State Physics (Tokyo) for hospitality when early stages of this work were underway.  P.C., P.C. and Y.K. acknowledge the Aspen Center for Physics and NSF Grant No. PHY-1607611 where this work was discussed and further developed.

\end{document}



\baselineskip24pt

\maketitle 

\newcommand{\nocontentsline}[1]{}

\setcounter{tocdepth}{2}
\tableofcontents




\clearpage

\section{M$\ddot{\bf{o}}$ssbauer Spectroscopy}
\subsection{M$\ddot{\text{o}}$ssbauer Spectrum: A Conceptual Discussion}
\Mb spectroscopy averages the environment of the nucleus over
time-scales of order a nanosecond, \red{the lifetime of the nuclear excited state}. 
In our experiments, quantum fluctuations in the charge of the f-electrons 
couple capacitively to the $^{174}$Yb nucleus, perturbing the nuclear
excitation energies.  
Provided the time-scale of these charge fluctuations are slower or comparable 
to a nanosecond, they modify the line-shape of a nuclear 
transition. 

\begin{figure}[h!]
\centering
\includegraphics[width=0.7\linewidth]{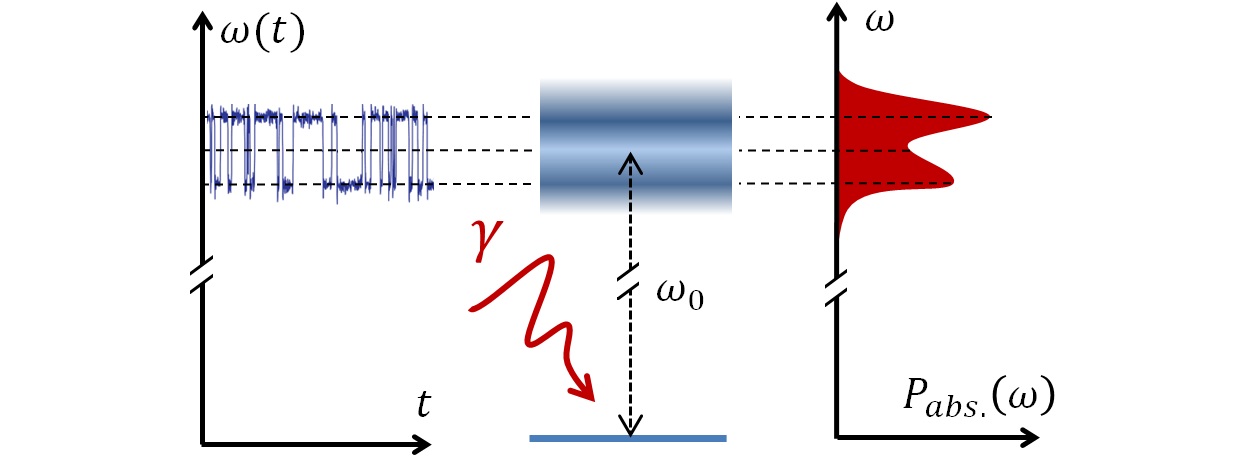}
\caption{\small Charge fluctuation in the environment capacitively couple to a two-level system (e.g. a nuclear transition) and fluctuate its resonate frequency in time. An absorption spectroscopy on this two-level leads to a modified lineshape $P(\omega)$ which provide information about the charge fluctuation spectrum $\chi''(\omega)/\omega$.}\label{fig:lineshape}
\end{figure}

The nuclear transition $\vert i\rangle \longrightarrow\vert f\rangle $ can be treated as
a two-level system 
with transition energy $E_{f}-E_{i}= \hbar
\omega_{0}$.   
The classic theory of lineshape broadening
 was developed by
Anderson, Kubo and Blume\cite{Anderson54,Kubo54,Blume68_1,Blume68_2},
before the advent of modern many-body theory. 
Their theory can be reformulated in a 
modern context\cite{Komijani16}, treating 
two level nuclear system as a 
single fermion with Hamiltonian $H_{0}= \hbar \omega_{0}
d\dg  d $. The energy shift of the nuclear energy level is 
then described by adding in an interaction $H_{I}= \alpha n_{f}
d\dg d$,  where  $\alpha$ is the monopole-like coupling to the 
occupancy  $\hat n_{f}$ of the f-orbitals. Thus a
static valence change of the f-state produces an energy shift $\alpha
$ in the \Mb absorption line.
With this formulation, the change in the \Mb absorption line-shape (Fig.\,\ref{fig:lineshape}) can
be described in terms of the self-energy $\Sigma _{d} (\omega)$ of the d-fermion.

The absorption cross-section\cite{Komijani16} 
at a Doppler shift energy $\hbar \omega$, 
$P (\omega+\omega_{0})$
is then related to the Green's function  of the d-fermion 
$G_{d} (\omega) = (\omega-\Sigma_{d} (\omega))^{-1}$
via 
\begin{equation}
\hspace{-.3cm}P(\omega+\omega_0)\propto {\rm Im }G_{d} (\omega-i\eta )=\frac{\Sigma''_d}{(\omega-\Sigma'_d)^2+\Sigma''^2_d}. 
\end{equation}
To leading order in perturbation
theory, 
\begin{equation}\label{}
\qquad \Sigma''_d (\omega) =\frac{\alpha^2T}{\omega}\chi''_\rho(\omega),
\end{equation}
where $\chi''_\rho(\omega)$ is the dynamical charge susceptibility of
the f-state at frequency $\omega$. In this way, \Mb spectroscopy
provides a direct probe of low-frequency valence fluctuations. 

The energy shift $\alpha\sim \hbar/\tau $ 
sets the characteristic averaging time-scale $\tau\sim $1ns over
which \Mb probes the surrounding charge fluctuations. 
If $\chi'' (\omega)/\omega$ is described by a Lorentzian
of width $\Lambda_{vf}$, then the resulting power-spectrum develops
a two peaked structure provided the charge fluctuation rate
$\Gamma_{vf}\ll \alpha $ is smaller than $\alpha $, but develops a
single peak-structure when $\Gamma_{vf}\gg \alpha $ is much faster
than the isomer shift $\alpha $.

\subsection{Importance in Studies of Heavy Electron Materials}
In heavy electron materials like \ybal, charge fluctuations are known
to lead to valence fluctuations \cite{Varma1976,sampathkumaran86}.
Here strong Coulomb repulsion within the f orbitals
restricts the partially filled f-shell occupancies to 
two orbital configurations, which in \ybal are the 4f$^{\
13}$(Yb$^{3+}$) and 4f$^{\ 14}$(Yb$^{2+}$).  Hybridization with the
conduction sea then causes valence fluctuations, Yb$^{2+}
\rightleftharpoons$Yb$^{3+}+ e^{-}$.  The simultaneous coexistence of
these hybridized configurations was first established by
core-level X-ray spectroscopy; the strength of the separate
absorption spectra of the two instantaneous configurations gives an
average Yb valence of about {2.7} in \ybal\cite{Okawa10}.  A series of
classic experiments in the late 1970s searched carefully for the
corresponding two valence configurations in \Mb studies
\cite{oldmoss,oldmoss2}, but they were never seen. Unlike 
static mixed valence known in chemistry (eg in iron oxide), the 
quantum mechanical valence
fluctuations of a heavy fermion system are homogeneous, corresponding
to hybridization energy scales of $10-1000$K that are too fast to be
observed by \Mb spectroscopy; usually the observed isomer shift 
is a single absorption feature, centered around the average of the two
valences. This is what makes the new observations so fascinating, 
for the line-splitting that develops, without a phase
transition, at low
temperatures appears to correspond to a slow-down of dynamic valence
fluctuations to the point where they become visible in the \Mb
spectroscopy.

\subsection{Synchrotron-radiation (SR)-based $^{174}$Yb M$\ddot{\text{o}}$ssbauer spectroscopy \label{a3}}

\begin{figure}
 \begin{center}
   \includegraphics [width=1.0\linewidth, angle=0, clip]  {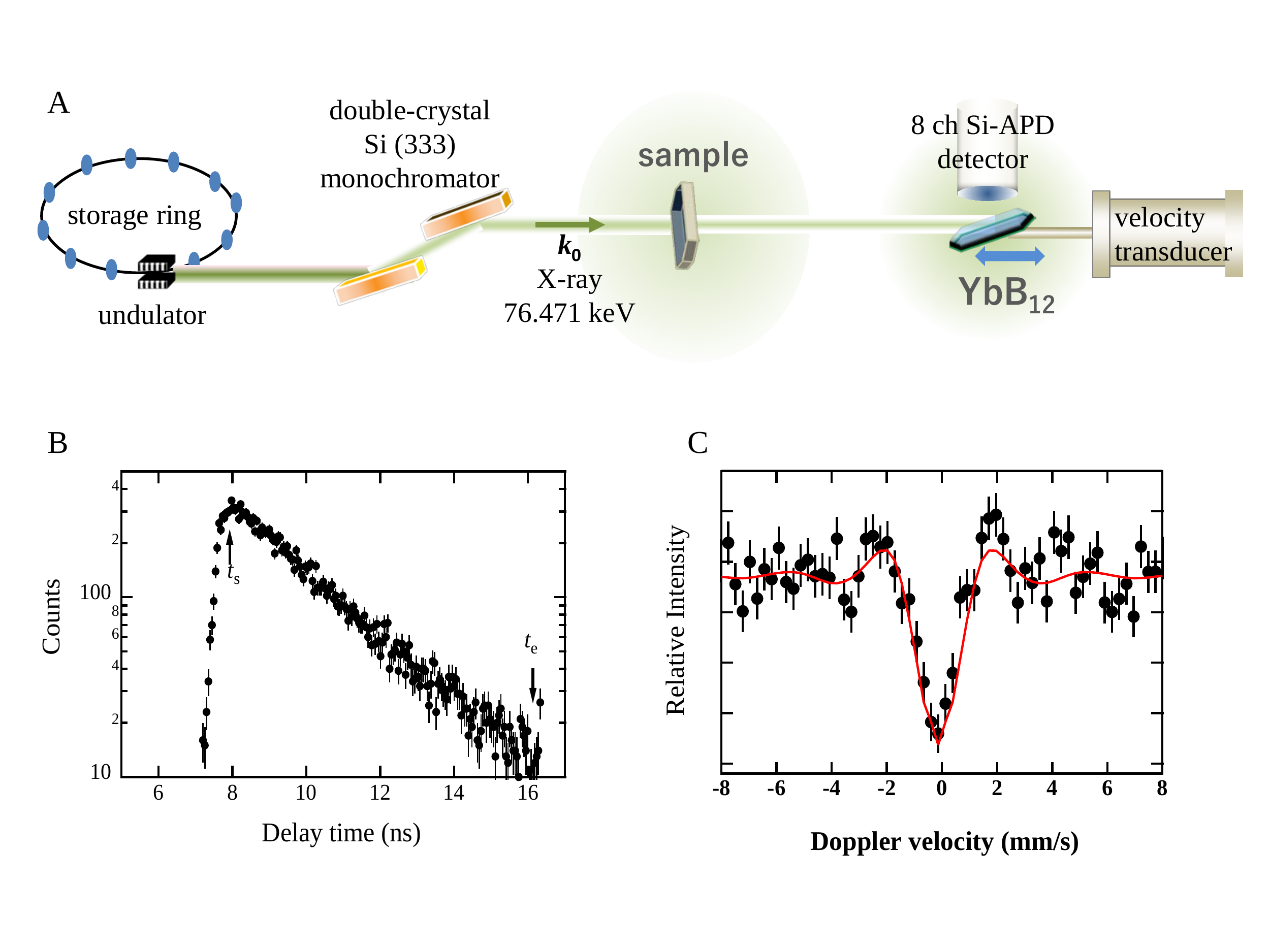}
   \caption{
(A) Schematic illustration of the experimental set up for the synchrotron-radiation-based $^{174}$Yb M$\ddot{\text{o}}$ssbauer spectroscopy. 
The $^{174}$Yb nuclear resonance was obtained by synchrotron radiation using a double-crystal Si (333) monochromator. 
A sample and the single-crystalline YbB$_{12}$ were cooled below 20K and at 26K in each cryostat, respectively. 
A multielement Si avalanche photodiode (APD) detector was used to measure delayed incoherent emission 
from $^{174}$Yb nuclei in the single-crystalline YbB$_{12}$ in the spatially incoherent scattering geometry. 
(B) Typical delay time spectrum of the single-crystalline YbB$_{12}$ at 26K. 
The closed circles with error bar indicate the observed spectrum. 
(C) Synchrotron-radiation-based $^{174}$Yb M$\ddot{\text{o}}$ssbauer spectrum 
of the powder YbB$_{12}$ sample at 20K. 
The delayed scattering signals were accumulated at each Doppler velocity with the time window from $t_{\rm s}$ to $t_{\rm e}$ as shown in (B). 
The closed circles with error bar and the red solid line present the observed and the analytical spectra, respectively.
}
\label{Fig:Fig_174YbMoss}
\end{center}
\end{figure}

The synchrotron-radiation-(SR)-based M$\ddot{\text{o}}$ssbauer spectroscopy is a new state-of-the-art experimental technique 
to investigate electronic states of M$\ddot{\text{o}}$ssbauer atoms in compounds \cite{Ruby74,Seto09}. 
Figure \ref{Fig:Fig_174YbMoss}(A) shows the experimental setup for the SR-based $^{174}$Yb  
($E_{\gamma} = 76.471$keV between the {$I_g =0$ and $I_e=2$} nuclear states)  
M$\ddot{\text{o}}$ssbauer spectroscopy at BL09XU and BL19LXU beamlines on SPring-8 \cite{Masuda14}. 
The SR pulse was monochromatized to approximately 5eV at $E_{\gamma} $ by using a double-crystal Si (333) monochromator 
and then encountered a sample including $^{174}$Yb nuclei in a cryostat. 
After passing through the sample, the monochromatized SR pulse encountered a single-crystalline YbB$_{12}$ (not enriched $^{174}$Yb) 
known as a scatterer. 
This YbB$_{12}$ scatterer was kept at 26K and moved in the triangle mode by a velocity transducer 
to create a relative Doppler velocity $\hbar \omega_{\rm D}$ between the sample and the scatterer. 
Note that very strong scattering due to electrons in matter occurs promptly after the SR pulse, 
but the scattering due to resonant nuclear excitation is delayed by the finite lifetime of the excited nuclear state.
Thus, we independently observe very weak nuclear resonant scattering in the time-domain measurement.
The scattering intensity of the monochromatized SR pulse passing through the sample in the forward direction was measured 
by these delayed resonant scattering signals from $^{174}$Yb nuclei in the YbB$_{12}$ scatterer after each prompt SR pulse 
using a multi-element Si avalanche photodiode (APD) detector in the spatially incoherent scattering geometry.

The delayed scattering signal $S (\hbar \omega_{\rm D}, \tau)$ after each SR prompt pulse in the nuclear recoil-free absorption process is given by 
\begin{equation}
S (\hbar \omega_{\rm D}, \tau)  \propto  \int_0^{z^{\rm sc}} dz \exp( - \mu_e^{\rm sc} z) 
\left| \int d\omega \frac{\Gamma_0/2 \exp(- i \hbar \omega \tau/\Gamma_0)}{\hbar (\omega - \omega_{\rm D}) - i \Gamma_0/2} 
A^{\rm sa}(\hbar \omega) A^{\rm sc} (\hbar \omega, \hbar \omega_{\rm D}, z) \right|^2,
\label{eqn:Sot}
\end{equation}
where
\[
A^{\rm sa}(\hbar \omega)  = A_0^{\rm sa} \exp \left( - i \frac{\mu_n^{\rm sa} z^{\rm sa} \Gamma_0/2 }{\hbar \omega - i \Gamma_0/2} \right) ,
\]
\[
A^{\rm sc}(\hbar \omega, \hbar \omega_{\rm D}, z)  = A_0^{\rm sc} \exp \left( - i \frac{\mu_n^{\rm sc} z \Gamma_0/2}{\hbar (\omega - \omega_D) - i \Gamma_0/2} \right) ,
\]
$\tau = t/\tau_0$ is a dimensionless time, $\tau_0 =\hbar / \Gamma_0$ the lifetime of the excited nuclear state, 
$A^{\rm sa}(\hbar \omega)$ $(A^{\rm sc}(\hbar \omega, \hbar \omega_{\rm D}, z) )$ represents the radiation amplitude in depth $z^{\rm sa}$ $(z)$ 
of the sample (scatterer), 
$z^{\rm sa}$  $(z^{\rm sc})$ and $\mu_n^{\rm sa}$ $(\mu_n^{\rm sc})$ denote the thickness and the nuclear resonant absorption coefficient 
of the sample (scatterer), respectively, and $\mu_e^{\rm sc}$ represents the electronic absorption coefficient of the scatterer \cite{Smirnov07}.
The delayed scattering signals $S (\hbar \omega_{\rm D}, \tau)$ were accumulated at each $\hbar \omega_{\rm D}$ within the time window 
from $\tau_{\rm s} = t_{\rm s}/\tau_0$ to $\tau_{\rm e} = t_{\rm e}/\tau_0$ to measure a SR-based M$\ddot{\text{o}}$ssbauer absorption spectrum 
in the energy-domain measurements.
Figure\,\ref{Fig:Fig_174YbMoss}(B) shows a delay time spectrum $S (\tau) (\propto \int S (\hbar \omega_{\rm D}, \tau) d\omega_{\rm D} )$ from $^{174}$Yb nuclei 
in the YbB$_{12}$ scatterer measured in our experiment. 
Since the lifetime of the $I_e=2$ excited state of the $^{174}$Yb nucleus is $\tau_0 = 2.58$ns, 
as seen in Fig.\,\ref{Fig:Fig_174YbMoss}(B), a typical time window in our experiments was in the range from $\tau_{\rm s} \sim 3.1$ ($t_{\rm s} \sim 8$ns) 
to $\tau_{\rm e} \sim 6.2$ ($t_{\rm e} \sim 16$ns) to measure SR-based $^{174}$Yb M$\ddot{\text{o}}$ssbauer spectra.

\subsection{Oscillatory Effects of the Finite-Time Window}

\red{
The intensity of $S (\hbar \omega_{\rm D}, \tau)$ around $|\hbar \omega_{\rm D}| \sim 0$ for the YbB$_{12}$ sample oscillates as a function of $\tau$ 
about the exponential curve as shown in Ref. \cite{Lynch60} for $^{57}$Fe M$\ddot{\text{o}}$ssbauer resonance 
where in general, $|\hbar \omega_{\rm D}| \sim 0$ means energy near the M$\ddot{\text{o}}$ssbauer transition of $^{174}$Yb nuclei in a sample.
These oscillations in $S (\hbar \omega_{\rm D}, \tau)$ are dynamical effects of scattering processes in the sample and the scatterer.
Since the frequencies of the oscillation components vary as $\hbar \omega_{\rm D}$, these oscillation behaviors in $S (\hbar \omega_{\rm D}, \tau)$ 
are smeared out in $S (\tau)$ as seen in Fig.\,\ref{Fig:Fig_174YbMoss}(B). 
Experimentally, the intensities at $\hbar \omega_{\rm D}$ in a spectrum, $S (\hbar \omega_{\rm D}) (= \int_{\tau_{\rm s}}^{\tau_{\rm e}} S (\hbar \omega_{\rm D}, \tau) d\tau)$, 
depend not only on $\hbar \omega_{\rm D}$ but also on the time window in the accumulation. 
Thus, a shape of an absorption component in a spectrum is affected by the time window in the accumulation and 
a Lorentz-function line shape with the expected width should be obtained in a spectrum using a thin-sample with the time window 
from $\tau_s \sim 0$ to $\tau_e \sim \infty$.
Figure\,\ref{Fig:Fig_174YbMoss}(C) shows the SR-based $^{174}$Yb M$\ddot{\text{o}}$ssbauer spectrum of the powder YbB$_{12}$ (not enriched $^{174}$Yb) sample at 20K. 
As seen in this figure, the oscillatory components were observed in background and the full-width at half-maximum of the main absorption component 
at 0 mm/s was evaluated to be 1.2 mm/s which is much narrower than that ($2\Gamma_0 = 2.00$ mm/s) 
expected from $\tau_0$ of the $I_e=2$ excited state of $^{174}$Yb nucleus.
These phenomena were observed in the $^{57}$Fe M$\ddot{\text{o}}$ssbauer spectra measured by the delayed coincidence method \cite{Hamillt68} 
and the SR-based $^{151}$Eu M$\ddot{\text{o}}$ssbauer spectroscopy \cite{Seto10}, 
indicating that these are related to the time-window effect in the accumulation of $S (\hbar \omega_{\rm D}, \tau)$. 
Thus, this time-window effect in the SR-based M$\ddot{\text{o}}$ssbauer spectroscopy is important for M$\ddot{\text{o}}$ssbauer resonances 
with short $\tau_0$s.
The observed SR-based $^{174}$Yb M$\ddot{\text{o}}$ssbauer 
spectrum of YbB$_{12}$ is well analyzed using one nuclear transition 
with this time-window effect and the effective thicknesses of the YbB$_{12}$ sample and the scatterer. 
Since $\tau_{\rm s} \sim 3.1$ is much larger compared to $\tau_{\rm s} \sim 0$, 
then the higher energy resolution than the conventional $^{174}$Yb M$\ddot{\text{o}}$ssbauer spectroscopy was achieved in our experiments.
The enhanced energy resolution may be essential for extracting time-dependent hyperfine interaction parameters in $\beta$-YbAlB$_4$.
}

\section{SR-based $^{174}$Yb M$\ddot{\text{o}}$ssbauer Spectroscopy of Single Crystal $\beta$-YbAlB$_4$}

\subsection{Experimental Specifics \label{b1}} 
High-purity single crystals of $\beta$-YbAlB$_4$ were grown by the Al-flux method \cite{Macaluso07}.  
Energy-dispersive X-ray analysis indicated no impurity phases, no inhomogeneities and a ratio Yb:Al of 1:1 within the detection limits of the equipment.  
The SR-based $^{174}$Yb M$\ddot{\text{o}}$ssbauer experiments were carried out at the BL09XU and BL19LXU beamlines on SPring-8 
(see Fig.\,1 A of the paper) \cite{Masuda14}, using SR to excite the $E_{\gamma}=$ 76.471keV M$\ddot{\text{o}}$ssbauer transition between
the $I_g=0$ ground and $I_e=2$ excited states of the $^{174}$Yb nuclei contained in the compound. 
The energy level diagram of the excited $^{174}$Yb nuclear state is sensitive to the surrounding charge configuration and provides real-time information 
about the Yb valence (see Fig.\,1 C of the paper). Measurements were carried out under both ambient and external pressure,  
and the $c$-axis of the single crystalline $\beta$-YbAlB$_4$ samples was aligned along the propagation vector $\Vec{k}_0$ of the incident X-ray. 
{Under external pressure, the single crystalline samples were loaded into a sample cavity of the Inconel alloy gasket in a clamp-type diamond-anvil cell (DAC) 
with ruby crystals and mixtures of methanol-ethanol or Daphne7474 as a pressure-transmitting medium to ensure hydrostatic conditions at 2K.
Pressure was calibrated by measuring the wavelength shift of the $R_{1}$ luminescence line of the ruby crystals in the clamp-type DAC.
Since the sizes of the single crystalline $\beta$-YbAlB$_4$ samples in the sample cavities were comparable with that of the incident X-ray beam, 
the collimators were used to obtain $^{174}$Yb M$\ddot{\text{o}}$ssbauer spectra of satisfactory quality.}
The high intensity of the incident  X-ray made it possible to carry out these experiments without isotopic purification.

\red{Our measurements were performed using cryostat with liquid Helium bath (SM4000 Oxford Instrument). The vibration of the sample/DAC can be ignored within out velocity scale for $^{\rm 174}$Yb M\"ossbauer spectroscopy.
The pressure medium for the diamond-anvil compression was a methanol-ethanol mixture that works best up to about 10 GPa.
In liquid media and 
in spectroscopic and diffraction measurements, this is the preferred
medium for diamond-anvil compression. In resistivity
measurements, the wire contacts  are 
connected to the sample using conductive paste, which unfortunately
reacts with alcohol, so methanol-ethanol mixtures are not used in
high-pressure resistivity measurements. 
Thus there can be sometimes be apparent discrepances between pressure-dependent
transport and spectroscopy
measurements that arise due to the different
pressure-transmitting media involved.
}

\subsection{M$\ddot{o}$ssbauer  Absorption Line-Shapes \label{b2}}

In the $^{174}$Yb M$\ddot{\text{o}}$ssbauer effect, the resonance is an electrical quadrupole $E2$ transition 
from the $I_g =0$ ground to $I_e=2$ excited nuclear states. 
The five nuclear transitions with $\Delta I^z (= I_g^z - I_e^z) = 0$, $\pm 1$, and $\pm 2$ are allowed 
between the ground and excited nuclear states. 
In no electrical quadrupole and magnetic hyperfine interaction, these five allowed transitions have the same energy 
and then we observe an absorption component in a M$\ddot{\text{o}}$ssbauer spectrum, as seen in Fig. \ref{Fig:Fig_174YbMoss}(C). 
The pure electrical quadrupole interaction splits the $I_e=2$ excited nuclear state into three sublevels with $I_e^z = 0$, $\pm 1$, and $\pm 2$ 
within an axial symmetry by 
\begin{equation}
\mathcal{H}_Q  =   \frac{1}{4} \frac{eQ V_{zz}}{I_e(2 I_e - 1)}  \left[3 (I_e^z)^2 - I_e (I_e + 1) \right], 
\label{eqn:HF_eq02}
\end{equation}
where the second derivative of the potential $V_{zz} = \partial ^2 V/\partial z^2$ is the $z$ component 
of the diagonalized electric-field gradient (EFG) tensor and $Q$ is the nuclear quadrupole moment of the $I_e=2$ excited nuclear state of the $^{174}$Yb nucleus.
Thus, three allowed nuclear transitions with $\Delta I^z = 0$, $\pm 1$, and $\pm 2$ have different energies. 
In the nuclear resonant scattering using SR, the polarized SR provides the possibility of selective excitations in allowed nuclear transitions. 
Due to the perfect $\sigma$-polarization of the incident X-ray in our experiments, the excitation probabilities in these $^{174}$Yb nuclear transitions 
are very sensitive to the direction of the $z$-axis and the symmetry of the diagonalized EFG tensor at the Yb site \cite{Hannon88}. 

For $\beta$-YbAlB$_4$ with the orthorhombic $Cmmm$ structure \cite{Macaluso07}, the principal axes of the diagonalized EFG tensor 
at one crystallographic Yb site are along the three crystal axes because the local symmetry at the Yb site is $m2m$. 
Since the Yb ions are sandwiched between boron layers and are centered between seven-member boron rings, 
the principal $z$-axis of the diagonalized EFG tensor at the Yb site is parallel to the $c$-axis of $\beta$-YbAlB$_4$, 
which is the quantization axis $\Vec{z}_J$ of the hyperfine interactions at the $^{174}$Yb nuclear in $\beta$-YbAlB$_4$.
In our experimental conditions where $\Vec{z}_J$ at the Yb site in $\beta$-YbAlB$_4$ was parallel to $\Vec{k}_0$,  
the $I_g = 0 \rightarrow  I_e^z = \pm 1$  nuclear transitions are only selected in the five $E2$ nuclear transitions 
of the $^{174}$Yb M$\ddot{\text{o}}$ssbauer resonance \cite{Hannon88}, 
which have the same energy within an assumption of the axial symmetric EFG tensor.

The possibilities of selective excitations in allowed nuclear transitions and the higher energy resolution are crucial to discuss dynamics of valence fluctuations 
and the electronic states of the Yb ions in $\beta$-YbAlB$_4$ via the SR-based $^{174}$Yb M$\ddot{\text{o}}$ssbauer spectroscopy.

As seen in Figs.\,2 A and B of the paper, the broadenings of two nuclear transitions are observed in the SR-based $^{174}$Yb M$\ddot{\text{o}}$ssbauer spectra 
for $\beta$-YbAlB$_4$, which is attributed to relaxation between two nuclear transitions. 
In the M$\ddot{\text{o}}$ssbauer resonances, a relaxation occurs in the same $\Delta I^z$ nuclear transitions with different energies 
caused by time-dependent hyperfine interactions.
The M$\ddot{\text{o}}$ssbauer absorption line shapes in the presence of the time-dependent hyperfine interactions have been discussed 
in the stochastic ways \cite{Anderson54,Kubo54,Blume68_1,Blume68_2}.
In the case of slow relaxation, $1/\tau_f < 2 \varepsilon$, where $2 \varepsilon$ represents a energy difference 
between these two nuclear transitions without the relaxation effects and $\tau_f$ denotes the characteristic timescale 
in the time-dependent hyperfine interactions, 
the M$\ddot{\text{o}}$ssbauer absorption line shape for two nuclear transitions is expressed in the stochastic model as follow:
\begin{equation}
I(\hbar \omega_{\rm D})  \propto 
\left( (1-R) \frac{\Gamma_0/2 + W +(1/x)(\hbar \omega_{\rm D} + Wx)}{(\hbar \omega_{\rm D} + Wx)^2 + (\Gamma_0/2 + W)^2} + 
 (1+R) \frac{\Gamma_0/2 + W - (1/x)(\hbar \omega_{\rm D} - Wx)}{(\hbar \omega_{\rm D} - Wx)^2 + (\Gamma_0/2 + W)^2}  \right),
\label{eqn:st_slow}
\end{equation}
where $2 W = 1/\tau_f$, $x =  [(\varepsilon/W)^2 -1]^{1/2}$, and $R$ represents small difference in intensity between two nuclear transitions.

\red{In order to combine Eq.\,\ref{eqn:st_slow} with Eq.\,\ref{eqn:Sot} both amplitude and phase information of $A^sa(\hbar\omega)$ are required. The amplitude $I(\hbar\omega)\sim |A^{sa}(\hbar\omega)|^2$ is given by Eq.\,\ref{eqn:st_slow}. To find the phase, we have fitted Eq.\,\ref{eqn:st_slow} with two Lorentzians which can be easily continued to the complex plane.}

Thus the relaxation brings about a broadening and distortions as well as a shift of the absorption lines. 
For $1/\tau_f (= 2 W) \ll  2 \varepsilon$, $I(\hbar \omega_{\rm D})$ is the same as two Lorentz-functions centered at 
$\hbar \omega_{\rm D} = \pm  Wx \sim \pm \varepsilon$, which is expected in the absence of any fluctuations in the hyperfine interactions.
In the case of fast relaxation, $1/\tau_f > 2 \varepsilon$, the M$\ddot{\text{o}}$ssbauer absorption line shape is given by 
\begin{equation}
I(\hbar \omega_{\rm D})  \propto \left( \frac{(1+1/y)( \Gamma_0/2 + W(1-y))}{(\hbar \omega_{\rm D})^2 + (\Gamma_0/2 + W(1-y))^2} + 
 \frac{(1-1/y)(\Gamma_0/2 + W(1+y))}{(\hbar \omega_{\rm D})^2 + (\Gamma_0/2 + W(1+y))^2}  \right),
\label{eqn:st_fast}
\end{equation}
where $y =  [1 - (\varepsilon/W)^2]^{1/2}$.
For $1/\tau_f (= 2W) \gg  2 \varepsilon$, $I(\hbar \omega_{\rm D})$ represents a Lorentz-function at $\hbar \omega_{\rm D} = 0$ 
(the first term in eq. (4)).
The fluctuations are so fast that the nucleus only feel an average hyperfine interactions. 
The collapse of a two-peaked structure related to two nuclear transitions occurs in a M$\ddot{\text{o}}$ssbauer spectrum  
as $2 \varepsilon \cdot \tau_f$ is comparable with and then becomes smaller than unity. 
Accordingly, the two-peaked structure observed in the M$\ddot{\text{o}}$ssbauer spectrum manifests slow relaxations 
in time-dependent hyperfine interactions.

\subsection{
Spectral Fits and  $^{174}$Yb Hyperfine Hnteractions in \ybal
\label{b3}
}

\subsubsection{Ambient pressure spectra}

\begin{figure}
 \begin{center}
  \includegraphics [width=0.7\linewidth]  {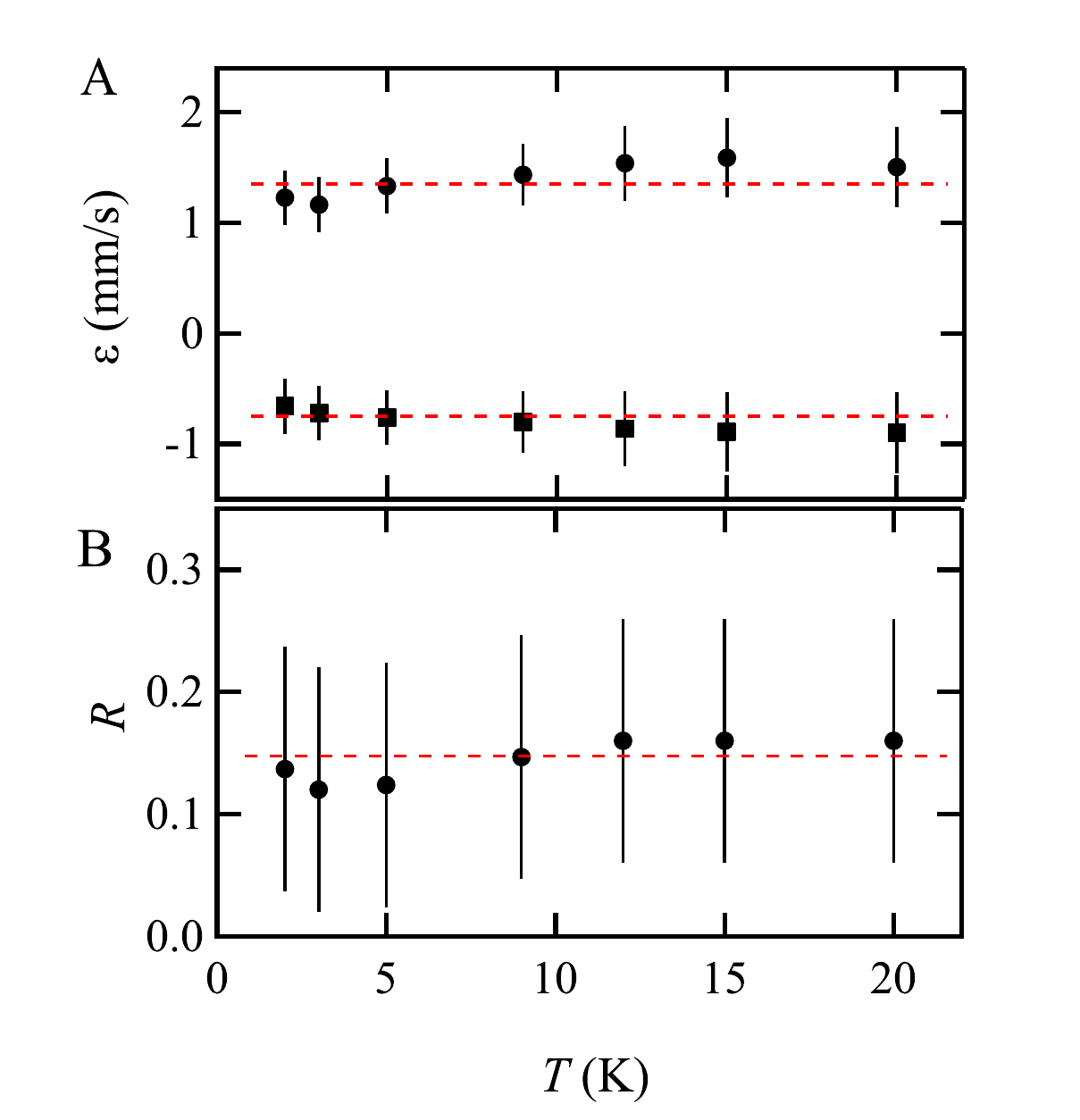}
\vspace*{+2.4cm}
   \caption{
(A) Refined energies $\varepsilon$ for two selected $I_g = 0 \rightarrow I_e^z = \pm 1$ nuclear transitions 
and (B) difference in intensity $R$ between these two nuclear transitions in $\beta$-YbAlB$_4$ as functions of temperature. 
The red broken lines are visual guides.
}
\label{Fig:Fig_EvsT}
\end{center}
\end{figure}

 The fitting procedures of the SR-based $^{174}$Yb M$\ddot{\text{o}}$ssbauer spectra for $\beta$-YbAlB$_4$ were performed using the equations 
in Section \ref{a3} with two selected nuclear transitions, 
including the effective thicknesses of the single crystalline $\beta$-YbAlB$_4$ samples and the YbB$_{12}$ scatterer 
and the time-window effect in the time-domain measurements in each experiment (see Section \ref{b1}). 
Figure\,\ref{Fig:Fig_EvsT}(A) shows the refined energies $\varepsilon$ for two $I_g = 0 \rightarrow I_e^z = \pm 1$ nuclear transitions 
in $\beta$-YbAlB$_4$ at ambient pressure.
As shown in this figure, the energy difference between these nuclear transitions is almost independent of temperature up to 20K. 

M$\ddot{\text{o}}$ssbauer spectroscopy has been used to discuss electronic states of atoms included M$\ddot{\text{o}}$ssbauer isotopes 
via the nuclear hyperfine interactions.
Although there are three main electrical and magnetic hyperfine interactions to determine the energies and the eigenstates of 
the ground and the excited nuclear states in a M$\ddot{\text{o}}$ssbauer isotope, 
the electrical monopole and electrical quadrupole interactions are important to discuss the electronic state of the atoms in a paramagnetic state.
In the experiments using the single crystalline $\beta$-YbAlB$_4$ samples, the degenerate $I_g = 0 \rightarrow  I_e^z = \pm 1$ nuclear transitions 
are only selected in the five $E2$ nuclear transitions. 
In a paramagnetic state with no external magnetic field, the energy of these degenerate nuclear transitions depends on 
both the electrical monopole and quadrupole interactions. 
We do not evaluate the electrical monopole and quadrupole interactions for two selected nuclear transitions in $\beta$-YbAlB$_4$ only 
from the spectrum observed in the experimental condition with $\Vec{z}_J \parallel \Vec{k}_0$. 
Thus, we have measured the SR-based $^{174}$Yb M$\ddot{\text{o}}$ssbauer spectrum at 5K in another experimental condition 
to evaluate the electrical monopole and quadrupole interactions for these two nuclear transitions. 
In the experiment, the $c$-axis of the single crystalline $\beta$-YbAlB$_4$ samples was tilted though 10 deg from the incident X-ray in the horizontal plane, 
that is, the angle $\theta$  between $\Vec{z}_J$ and $\Vec{k}_0$ was 10 deg. 

As shown in Fig.\,\ref{Fig:Fig_Moss10}, the characteristic features in the SR-based $^{174}$Yb M$\ddot{\text{o}}$ssbauer spectrum observed at 5K with $\theta$ = 10 deg 
are different from those in the spectrum at 5K with $\theta$ = 0 in Fig.\,2 A of the paper, due to the difference of the excitation probabilities 
in five allowed nuclear transitions between these experimental conditions. 
The excitation probabilities in these nuclear transitions were evaluated with $\theta$ = 10 deg. 
We have analyzed the observed SR-based $^{174}$Yb M$\ddot{\text{o}}$ssbauer spectrum at 5K 
using the evaluated excitation probabilities for the five $E2$ nuclear transitions. 
In this analysis, furthermore, the $\tau_f$ value and the energies for the degenerate $I_g = 0 \rightarrow  I_e^z = \pm 1$ nuclear transitions 
were fixed on those refined from the spectrum observed at 5K with $\theta$ = 0 within the experimental accuracies. 
The energy $\varepsilon$ of the $I_g = 0 \rightarrow  I_e^z = \pm 1$ nuclear transition is given by $\varepsilon = \delta - E_Q/2$ using 
the isomer shift $\delta$ and the electrical quadrupole splitting $E_Q$.  
As seen in Fig.\,\ref{Fig:Fig_Moss10}, the analytical spectrum agrees well with the observed one. 
The $\delta$ and $E_Q$ values were evaluated to be $\delta = 0.22$ mm/s and $E_Q = -2.3$ mm/s 
and $\delta = -0.76$ mm/s and $E_Q \sim 0$ mm/s for two selected nuclear transitions, respectively, by using the stochastic model. 

As shown in Fig.\,\ref{Fig:Fig_EvsT}(B), the intensity ratio $R$ between these two selected nuclear transitions is independent of temperature up to 20K 
within our experimental accuracy.
The selected nuclear transitions with $\delta = 0.27$ mm/s and $E_Q = -2.3$ mm/s and $\delta = -0.76$ mm/s and $E_Q \sim 0$ mm/s 
correspond to the Yb$^{3+}$ and Yb$^{2+}$ ionic states, respectively. 
Since all Yb sites are crystallographically equivalent in $\beta$-YbAlB$_4$, we assume the same recoil-free fraction (Lamb-M$\ddot{\text{o}}$ssbauer factor) 
of two selected nuclear transitions. 
The averaged Yb valence was evaluated to be 2.6(1) from the refined $R$ values, which is in good agreement with those (Yb$^{+2.75}$) obtainede
by hard X-ray photoelectron and X-ray absorption near edge structure spectroscopies \cite{Okawa10,Matsuda13}.

Since the nuclear transition for the Yb$^{2+}$ state shows a single line feature in the observed spectra, 
the contribution of the distribution of the ionic and electronic charges in the lattice to $E_Q$ is negligibly small at the Yb site in $\beta$-YbAlB$_4$ 
within our experimental energy resolution.  
The $E_Q$ value extracted for the Yb$^{3+}$ state thus provides information about the crystalline-electric-field (CEF) ground state of 
the Yb$^{3+}$ $4f$ electrons in $\beta$-YbAlB$_4$.
Since the degenerate $I_g = 0 \rightarrow I_e^z = \pm 1$ nuclear transitions are selected in the experimental condition with $\Vec{z}_J \parallel \Vec{k}_0$, 
the assumption of the axial symmetric EFG is reasonable within our experimental energy resolution.  
The $E_Q^{4f}$ due to the interaction between the nuclear quadrupole moment $Q$ and the quadrupole moment of the CEF ground state of rare-earth $4f$
electrons is given by
\begin{equation}
E_Q^{4f}  = - \frac{1}{4} e^2 Q (1 - R _Q)  \langle J || \alpha || J \rangle \langle r^{-3} \rangle_{4f} \langle 3 J_z^2 - J(J+1)  \rangle ,
\label{eqn:Eq4f}
\end{equation}
where $\langle r^{-3} \rangle_{4f}$ is the averaged value of $r^{-3}$ for the $4f$ electrons in a given $J$ state for a rare-earth ion, 
$\langle J || \alpha || J \rangle$ represents a reduced matrix element defined by the operator equivalent and $R_Q$ is the shielding factor \cite{Ofer68}. 
The states of the Yb$^{3+}$ $4f$ electrons are split to four Kramers doublets by the CEF interaction. 
In the case where $J_z$ is conserved, the expected $E_Q^{4f}$ values are evaluated to be $-14.7$, $-2.1$, $6.3$, and $10.5$ mm/s 
for the four Kramers doublets with $J_z = \pm7/2$, $\pm 5/2$, $\pm 3/2$, and $\pm 1/2$, respectively, 
using $\langle r^{-3} \rangle_{4f} = 13.83$ a.u., and $R_Q = 0.21(1)$ with $\langle J || \alpha || J \rangle = 2/63$ for $^2F_{7/2}$ and $Q= 2.12$ b \cite{Freeman62,Meyer79}. 
As the result of the extracted $E_Q$ value for the Yb$^{3+}$ state, the ground Kramers doublet for  the Yb$^{3+}$ $4f$ electrons 
is almost pure $J_z = \pm 5/2$ states in $\beta$-YbAlB$_4$ with a negligible admixture of the $J_z = \pm 1/2$ state.

\subsubsection{Finite pressure, low temperature spectra. }

As seen in Fig.\,2 B of the paper, the similar two-peaked structure was observed in the SR-based $^{174}$Yb M$\ddot{\text{o}}$ssbauer spectra 
at 0.25 and 0.7GPa at 2K.
It was confirmed by the SR X-ray diffraction measurements of $\beta$-YbAlB$_4$ under pressure \cite{Sakaguchi16} 
that the structural symmetry and the individual atomic coordination parameters do not change up to 3.5GPa at 7K.
Thus, the characteristic change in the features of the observed spectra around $\sim 1$GPa is not related to a pressure-induced structural transition 
in $\beta$-YbAlB$_4$ at 2K.
The observed spectra below 1.2GPa at 2K were analyzed by using the same stochastic model like that at ambient pressure 
 (see Section \ref{a3} and \ref{b2}).
As seen in Figs.\,2 A and B  of the paper and Fig.\,\ref{Fig:Fig_174YbMoss}(C), the absorption component observed 
in the SR-based $^{174}$Yb M$\ddot{\text{o}}$ssbauer spectrum at 2.3GPa and 2K is much narrower than that in the spectrum at 20 K 
and ambient pressure and comparable with that in the spectrum of YbB$_{12}$.
The observed spectrum at 2.3GPa and 2K was analyzed using a single absorption component  (see Section \ref{b2}).
Figure\,\ref{Fig:Fig_Evsp}(A) shows the refined energies $\varepsilon$ for two $I_g = 0 \rightarrow I_e^z = \pm 1$ nuclear transitions up to 1.2GPa at 2K 
and the refined absorption-peak position at 2.3GPa and 2K in $\beta$-YbAlB$_4$.
As shown in Fig.\,\ref{Fig:Fig_Evsp}(B), the intensity ratio $R$ between these two selected nuclear transitions is independent of pressure up to 1.2GPa at 2K 
within our experimental accuracy.
The absorption-peak position at 2.3GPa and 2K corresponds to the weighted average of two absorption-peak positions below 1.2GPa at 2K, 
revealing that the averaged electronic state of the Yb ions in $\beta$-YbAlB$_4$ is not changed by pressure up to $\sim$2.3GPa at 2K. 

Since the $^{174}$Yb M$\ddot{\text{o}}$ssbauer spectrum at 2.3GPa and 2K was well reconstructed by a Lorentz-function, 
$1/\tau_f \gg 2 \varepsilon$ in the stochastic relaxation model.
The width of the absorption component of $\beta$-YbAlB$_4$ sample was refined to be $\Gamma = 1.11(1)$mm/s at 2.3GPa and 2K, 
which is broader than $\Gamma_0 =$1.00mm/s expected from $\tau_0$ of the $I_e=2$ excited state of the $^{174}$Yb nucleus.
If we assume that this broadening comes from the relaxation between two nuclear transitions, $\Gamma$ is given by 
\[
\Gamma = \Gamma_0 + 2 W(1-y) \sim \Gamma_0 + \frac{\varepsilon ^2}{2} \tau_f
\]
from the first term of eq. (4).
The energy difference $2 \varepsilon$ between two $I_g = 0 \rightarrow I_e^z = \pm 1$ nuclear transitions was estimated to be 
$2 \varepsilon \sim 2.2(6) $ mm/s from the refined energies $\varepsilon$ for two $I_g = 0 \rightarrow I_e^z = \pm 1$ nuclear transitions 
for the Yb$^{3+}$ and Yb$^{2+}$ ionic states at ambient pressure and below 1.2GPa at 2K.
We have evaluated $\tau_f \sim 0.04 \tau_0 ~(\sim 0.1$ns) at 2.3GPa and 2K.
In M$\ddot{\text{o}}$ssbauer absorption experiments, the broadening from $\Gamma_0$ in the width of the Lorentz-function 
could occur due to the effective thickness of a M$\ddot{\text{o}}$ssbauer isotope in a sample.
Thus, the evaluated $\tau_f $ value is maximum one for $\beta$-YbAlB$_4$ at 2.3GPa and 2K.

\subsection{Statistical Analysis of the Spectra}
The entire focus of our paper involves the resonant \Mb
absorption spectra that lie at low Doppler velocity below $\pm$2mm/s.  In this region, the X-ray scattering off the nucleii is
unitary and the only source of error is the photon count. 
By contrast, the off-resonance \Mb signals at large
Doppler shift are dominated by multiple elastic scattering, 
leading to interference effects that give rise to systematic errors in
the data that are absent at low Doppler shift.  
 This leads to
modulations in background signal as a function of Doppler velocity that are related to the finite sample ($\beta$-YbAlB$_4$)
and absorber (YbB$_{12}$) thicknesses,  and the finite time window for photon counts\,\cite{Wolfgang,Seto10,Seto09,Smirnov07}. \\

Now we turn to the analysis of the resonant \Mb absorption. 
There are two aspects of the data
that we wish to highlight that 
clearly demonstrate a signal that rises well above the
noise level. There are two i

\subsubsection{Broadening of the Lineshape at low temperature}
A plot of FWHM of the spectral dip as a function of temperature and
pressure (Fig.\,\ref{Fig:Fig_3c_in}) clearly shows that the width of absorption
line broadens at low temperatures and pressures.  
This is in marked contrast with the fact that the spectral peaks
usually broaden at high temperatures. Furthermore, the evolution of
the FWHM vs.~pressure correlates with transition between a strange
metal to a Fermi liquid in previous transport experiments.

\subsubsection{Statistical significance of double-peak structure}
We have done a thorough statistical analysis of the data. The 
quality of the fit is dependent on the regression and total sum of
squares, given by 
\begin{equation}
SSR=\frac1N\sum_{i=1}^{N}(y_i-f_i)^2, \qquad SST=\frac1N\sum_{i=1}^{N}(y_i-\bar y)^2
\end{equation}
where $y_{i}$ is the data, $f_{i}$ is the fit to data, $\bar
y=\frac{1}{N}\sum_i y_{i}$ is the average of the data set and $N$ is
the number of data points. 
These variables are
used to compute the standard error via $\sigma_{tot}=\sqrt{SSR}$. However, it is more systematic to define two new parameters $R^2$ and adjusted $\bar R^2$ parameters:
\begin{equation}
R^2=1-\frac{SSR}{SST}, \qquad \bar R^2=1-\frac{N-1}{N-P-1}(1-R^2)
\end{equation}
where
the latter compensates for the number of fitting parameters, $P$.

Fig.\,\ref{fig4}(A) shows the attempt to fit the experimental
lineshape using a single Gaussian or Lorentzian peak. 
Such single-peak fits do not capture the data in either the 
central region or the tails,  as quantified by the low values of $R^2$
and $\bar R^2$. If, following Referee 3.  we disregard 
the key central region as noise, we can fit the tails using a Gaussian or
Lorentzian lineshape, providing a noise background against which we
can assess the probability that the splitting is a statistical
anomaly. This fit is shown in blue in Fig.\,\ref{fig4}(B). 

The blue baseline, based on the assertion that there is no splitting
allows us to now assess whether the observed deviation from a single
peak is statistically significant.  The error bars in the measurement
have been reliably established from the standard deviation in photon
counts. Taking the pessimistic assumption 
that each deviating point is the result of Gaussian noise about the
single peak background, we then have two points
with a standard deviation of 2$\sigma $ a central point with a
standard deviation of 3$\sigma$. The probability of this deviation is
then given by
\begin{equation}\label{}
{\cal P}= {\cal E} (2\sigma )^{2}{\cal E} (3\sigma ) \equiv  {\cal E} (4.84\sigma )
\end{equation}
where 
\begin{equation}\label{}
{\cal E} (x) = \frac{1}{\sqrt{2\pi\sigma }}\int_{x}^{\infty }e^{-\frac{x^{2}}{2\sigma^{2}}}.
\end{equation}
In other words, this data represents a 4.84$\sigma $ event, an event
that would occur less than one in a million experiments.  
This
analysis, combined with the other data that shows a systematic
broadening at low temperatures and low pressures establishes that our results
meet the gold standard (5$\sigma $)  for statistically significant
experimental observations. 

By contrast,  the double-peak fit to the data, a fit grounded in the
Anderson-Kubo formula for dynamical fluctuation, 
 is significantly more successful, as shown in Fig.\,\ref{fig4}(C). It
 is these fits that were used to extract the valence fluctuation rate $\tau_{f}^{-1}$.
Comparing the $R^2$ and $\bar R^2$ of these fits, we conclude that first the analytical Kubo-Anderson fit and next a Gaussian fit are the best fits to the data.

\subsection{Lamb-M$\ddot{\text{o}}$ssbauer factor on $^{174}$Yb nuclei in $\beta$-YbAlB$_4$}

The Lamb-M$\ddot{\text{o}}$ssbauer factor $f_{\rm LM}$ describing the recoil-free fraction of the nuclear absorption, 
is related to atomic motions for M$\ddot{\text{o}}$ssbauer isotopes in a matter. 
The $f_{\rm LM}$ is generally described by
\begin{equation}
f_{\rm LM} = \langle \exp[i (\Vec{k}_0 -\Vec{k}_f) \cdot \Vec{r}] \rangle 
= \exp[-\langle [\Vec{k}_0 \cdot \Vec{r}]^2 \rangle] = \exp[-k_0^2 \langle \Delta z^2 \rangle],\label{eq6} 
\end{equation}
where $\Delta z$ is an atomic displacement from a regular position in a matter along the direction of $\Vec{k}_0$. 
In scattering experiments, this is the effects of the zero point and temperature motions of the atom bound in a matter.
The temperature dependence of $f_{\rm LM}$ is 
\begin{equation}
f_{\rm LM} (T) = \exp \left( -E_{\rm R} \int_0^{\infty} \frac{F (E)}{E} \coth\frac{\beta E}{2} dE \right)  
\end{equation}
using the local (partial) phonon density of states $F (E)$ for a M$\ddot{\text{o}}$ssbauer atom, 
where $\beta = \frac{1}{k_{\rm B} T}$, $k_{\rm B}$ is Boltzmann constant and $E_{\rm R} (= \frac{E_{\gamma}^2}{2 m c^2})$ recoil energy.
In the Debye approximation for $\nu(E)$, $f_{\rm LM} (T)$ becomes
\begin{equation}
f_{\rm LM} (T) = \exp \left( -\frac{6E_{\rm R}}{k_{\rm B} \Theta_{\rm D} } \left[ \frac{1}{4} + \left(\frac{T}{\Theta_{\rm D}}\right)^2 \int_0^{\Theta_{\rm D} /T} \frac{x}{e^x-1} dx \right] \right).
\end{equation}
In the case of $T \ll \Theta_{\rm D}$, $f_{\rm LM} (T)$ has a formula of
\begin{equation}
f_{\rm LM} (T) = \exp \left( -\frac{E_{\rm R}}{k_{\rm B} \Theta_{\rm D}} \left[ \frac{3}{2} +\left(\frac{\pi}{\Theta_{\rm D}}\right)^2 T^2 \right] \right).\label{eq9}
\end{equation}

In SR-based M$\ddot{\text{o}}$ssbauer spectroscopy, the relative intensity $I$ integrated in the energy region of the spectrum 
with absorption components for a sample is proportional to the number and $f_{\rm LM}$ of M$\ddot{\text{o}}$ssbauer resonant nuclei in the sample. 
The same single crystalline $\beta$-YbAlB$_4$ sample was used in our SR-based $^{174}$Yb M$\ddot{\text{o}}$ssbauer experiments at ambient pressure. 
The size of this sample was much larger than that of the incident X-ray beam, indicating that the number of $^{174}$Yb nuclei in the beam was same in our experiments.
We have evaluated $I$ proportional to $f_{\rm LM} (T)$ from the observed SR-based $^{174}$Yb M$\ddot{\text{o}}$ssbauer spectra at ambient pressure 
with correction of scattering  conditions including the time-window effects.

In our experiments under pressure at 2K, as seen in Fig.\,2 B of the paper, $^{174}$Yb M$\ddot{\text{o}}$ssbauer spectra of satisfactory quality were obtained 
by using the collimators to absorb incident X-ray photons without passing though the single crystalline $\beta$-YbAlB$_4$ sample in the clamp-type DAC, 
which were made by Re and Pb metals with 250 and 800$\mu$m thickness, respectively.
Although the thicknesses of these samples were same within our experimental accuracies, 
the transmission intensities of the incident X-ray photons with $E_{\gamma}$ were slightly different though these collimators.
We have corrected these absorption effects to evaluate $I$ proportional to $f_{\rm LM} (p)$ 
from the observed SR-based $^{174}$Yb M$\ddot{\text{o}}$ssbauer spectra.

\section{Reproducibility after Thermal Cycling}
\begin{figure}[h!]
\centering
\includegraphics[width=0.6\linewidth]{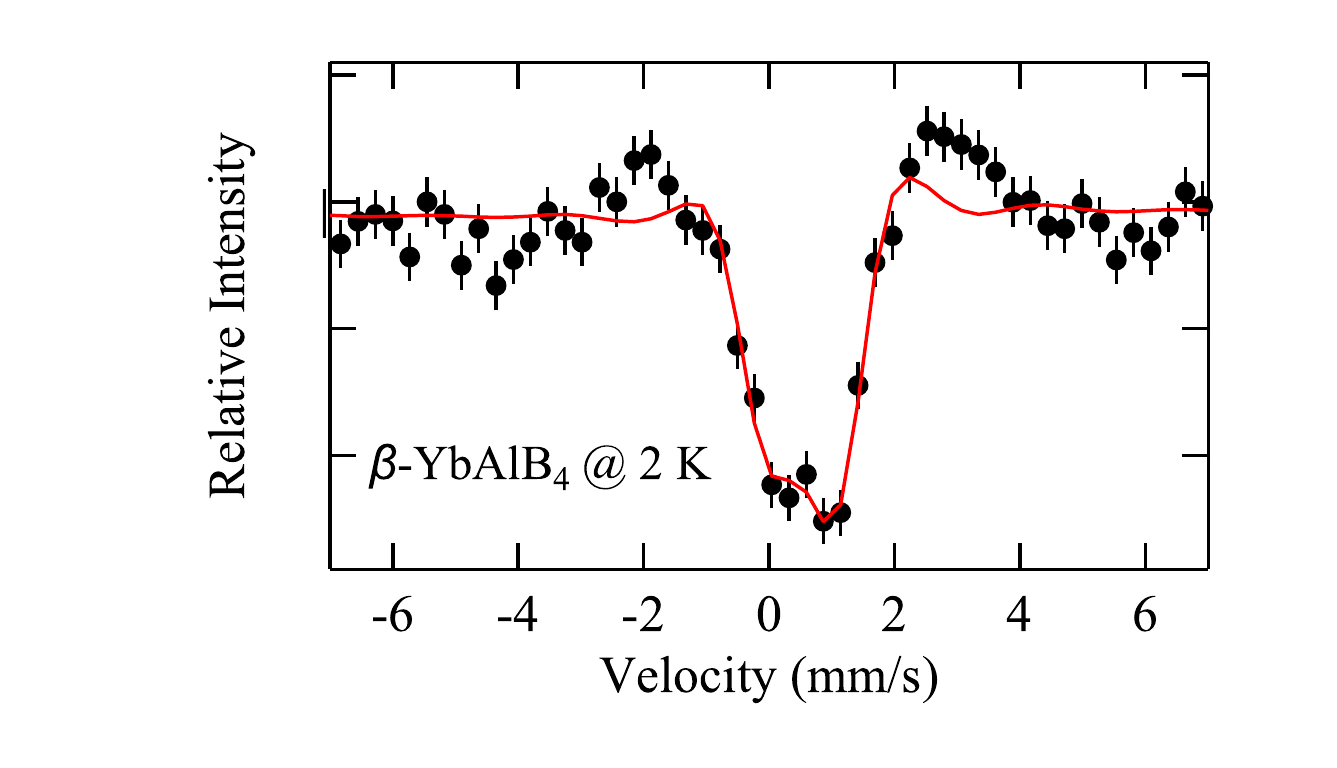}
\caption{\small M\"ossbauer spectra of the same $\beta$-YbAlB$_4$
sample studied in the paper after a thermal cycling at 2K under
ambient pressure to confirm the reproducibility of the splitting. For
comparison, the analytical curve in red is the same fit as that used
in the 2K ambient pressure data in the paper (Fig 2A in paper).}\label{figreproduce}
\end{figure}
Fig.\,\ref{figreproduce} shows the result of M\"ossbauer spectroscopy of the the same $\beta$-YbAlB$_4$ sample studied in the paper in a different thermal cycle at 2K and under ambient pressure. Although the details have changed the characteristic double-peak feature persists.

\section{Recent Characterization of Samples using Bulk Probes}
\begin{figure}[h!]
\includegraphics[width=\linewidth]{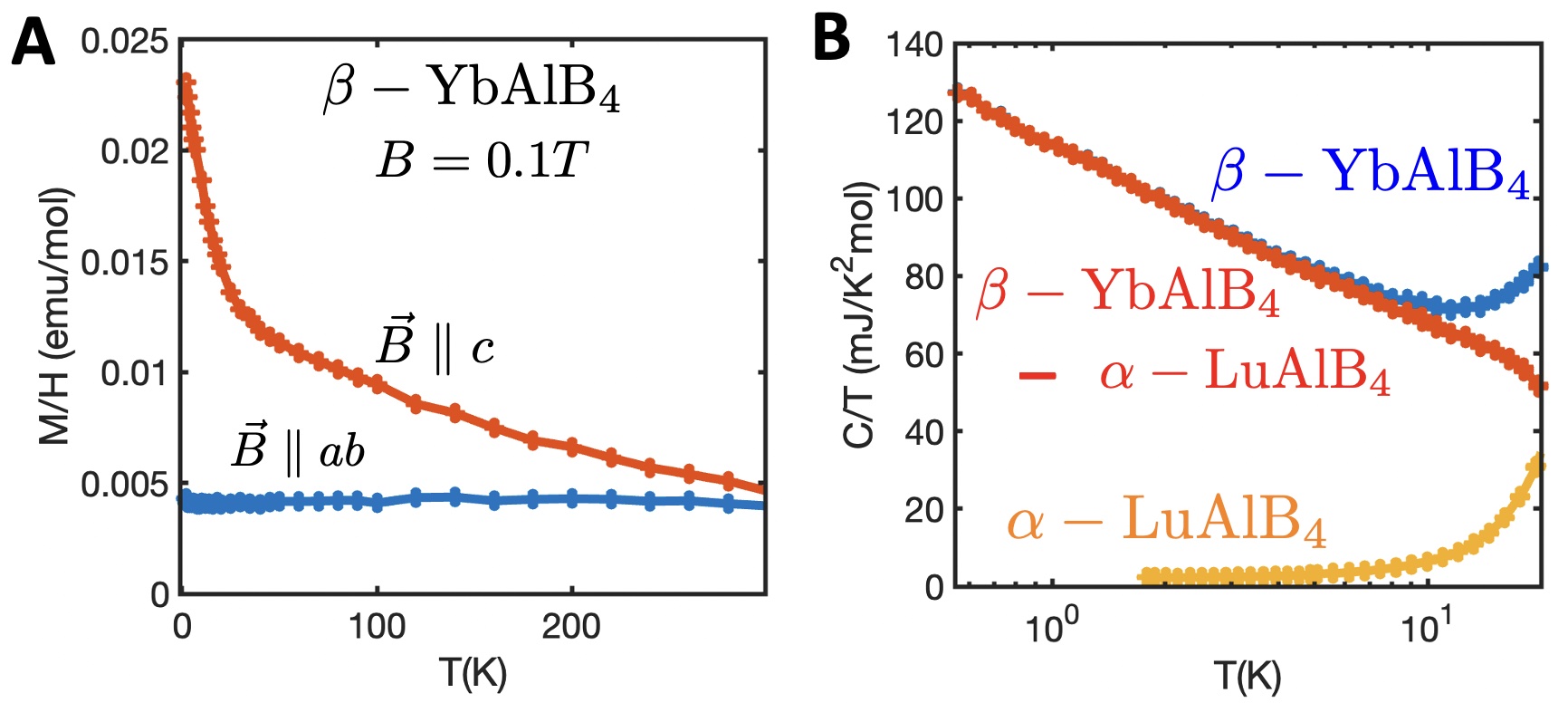}
\caption{\small (A) Magnetic susceptibility vs. temperature for field parallel to ab-plane and c-axis. (B) Heat capacity of \ybal{} and $\alpha$-LuAlB$_4$. The latter is subtracted from the former to get the electronic contribution to the $C/T$ shown in red.}\label{fignewthermo}
\end{figure}

\begin{figure}[h!]
\centering
\includegraphics[width=0.6\linewidth]{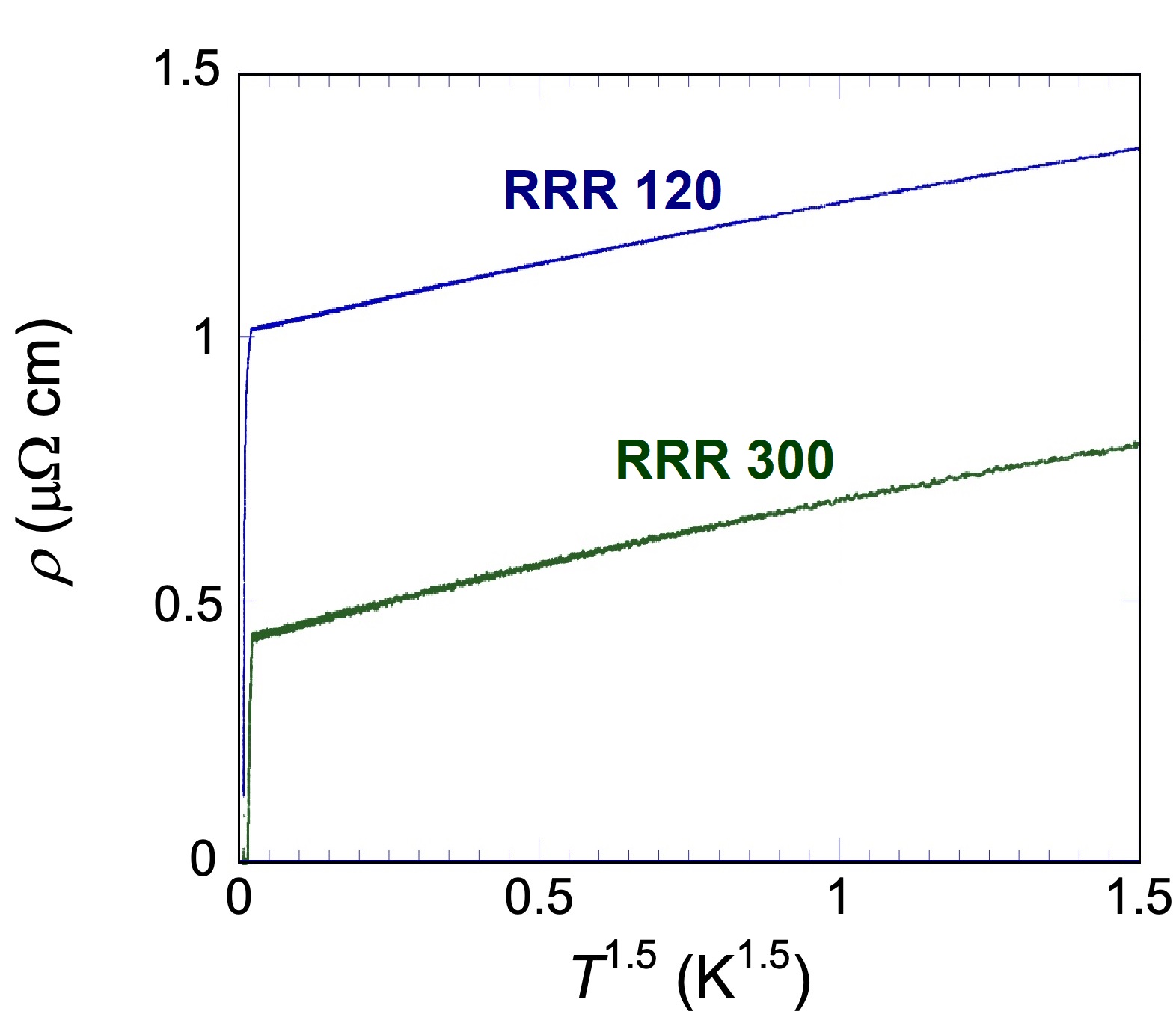}
\caption{\small Resistivity of $\beta$-YbAlB$_4$ as a function of temperature for two different residual resistivity ratios (RRRs). Note the sharpness of the superconducting transition for the selected single crystals with RRR$>$100 used in our M\"ossbauer experiments.}\label{figsctransition}
\end{figure}

In this section,  we briefly present new thermodynamical data taken from recently grown \ybal{ }crystals, similar to the ones used for the \Mb spectroscopy. As seen in Fig.\,\ref{fignewthermo}, the absence of any thermodynamic transition in $M/H$ and $C/T$ support the interpretation of the \Mb data as discussed in the next section. Fig.\,\ref{figsctransition} indicates the sharpness of superconducting transition for the same samples.


\section{Excluded Possibilities for Observed Results}

In the following we discuss a series of cross-checks to confirm the overall 
consistency of the results and that support of the charge fluctuation 
interpretation of the data.

\begin{itemize}

\item {{\bf Nuclear Origins.} The quantization axis $\Vec{z}_J$ of the diagonalized electric-field gradient (EFG) tensor 
at the Yb site with the $m2m$ local symmetry is parallel to the $c$-axis of $\beta$-YbAlB$_4$ with an orthorhombic $Cmmm$ structure, 
indicating $\Vec{z}_J \parallel \Vec{k}_0$ in our experiments. 
As a result of the angular momentum selection rule, only a single degenerate nuclear transition $I_g=0 \rightarrow  I_e^z=\pm 1$ is probed out 
of the five $E2$ nuclear transitions ($\Delta I^z  = 0, \pm 1,$ and $\pm 2$) of the $^{174}$Yb M$\ddot{\text{o}}$ssbauer resonance 
 (see Section \ref{b2}) \cite{Hannon88}.}

\item {{\bf Magnetic Origins}
The absence of long-range magnetic order in $\beta$-YbAlB$_4$ \cite{Matsumoto11,Tomita15} rules out magnetic hyperfine interactions 
as the origin of this splitting.
Generally, the electrical monopole interaction in the hyperfine interaction is related to a valence state of a rare-earth ion.}

\item{{\bf Disorder.}
All Yb sites are crystallographically equivalent in $\beta$-YbAlB$_4$ and it is known, 
from the SR X-ray diffraction measurements \cite{Sakaguchi16}, that the structural $Cmmm$ symmetry 
and the individual atomic coordination parameters do not change up to 3.5GPa at 7K.
Moreover, the residual resitivity ratio RRR more than 100 reveals that there is no quenched disorder in this material at low temperature.} 

\item{{\bf Antiferroquadrupolar (AFQ) Order.}
From the M\"ossbauer spectra alone, it is not possible to exclude
the scenario of an AFQ configuration with
staggered electric field gradients. However such a state of long-range 
order can  be directly ruled out from other observations. 
Such quadrupolar transitions are well
known in other heavy fermion systems 
(e.g.  UPd$_{3}$\cite{UPd3} and
CeB$_{6}$\cite{ceb6}) 
where they lead to 
a singularity in the specific heat and a cusp in the resistivity.
Neither features are observed. Finally,
there is no signature of any quadrupolar structural change in the high
resolution X-ray diffraction patterns.}

\item{{\bf Charge Density Wave Scenario}
There are two good reasons why a charge density wave (CDW) scenario
can be ruled out as an explanation for the observed effects we present here:

First, there is no evidence in $\beta$-YbAlB$_4$ for any phase transition into a
partially gapped phase in either transport or thermodynamic measurements (Figs.\,\ref{fignewthermo},\ref{figsctransition}), 
which all display smooth evolution as a function of decreasing temperature
with no observable discontinuities or other features.

More generally, we can rule out any ordered frozen configurations, which would have produced a
thermodynamic signal. The alternative, disordered frozen configurations would play
the role of quenched disorder, a possibility that is ruled out by
the tiny residual resistivity of less than 1$\mu\Omega$cm in the $\beta$-YbAlB$_4$ samples used in the M\"ossbauer experiments.

Second, the observed isomer shift in the 
\Mb spectra is inconsistent with a CDW.  
Isomer shifts result from small modifications of
nuclear excitation spectra 
induced by changes in the occupancies of orbitals with significant atomic
penetration.  3d orbitals of transition metals and 4f-orbitals of
rare earth atoms 
that 
penetrate deep into
an atom, ``communicating'' their valence to the nucleus via their
Coulomb-coupling to the innermost s-orbitals. 
The observed isomer shifts for $^{174}$Yb M\"ossbauer
isotopes,  
Yb$^{2+}$ and  Yb$^{3+}$ ions  are separated by about $1$mm/s
[see W. Henning, $et$ $al$. Z. Phys. {\bf 241} 138 (1971)], and 
the splitting observed to develop between the two peaks in the
spectrum of $\beta$-YbAlB$_4$ at 2 K is of the same order.
A CDW would typically occur in a conduction band orbital that
does not penetrate far inside the atom and thus does not 
influence the nuclear spectrum; moreover, a CDW typically
involves a fractional modulation in charge, far smaller than the integral
charge required to produce known isomer shifts.}

\item{{\bf Relaxational Broadened Static Double Peak}
Finally, one may envision a relaxational broadening mechanism to account for the discrepancy of the static fit to the lineshape. An electronic relaxation broadening mechanism, produced for
example by a phonon-induced change in the Yb environment would have
to be slow on a \Mb time scale to broaden the peaks. 
\Mb spectroscopy averages the environment of the nucleus over
time-scales of order a nano-second. Anything faster than this scale
simply leads to a single, motionally narrowed absorption line. 
To avoid this fate, a relaxational excitation would have to operate at 
at the micro-volt scale, a scale that is orders of magnitude smaller
that the characteristic scale of collective electronic or lattice excitations.}

\item{{\bf Mixture of Nuclear Transitions.}  
By varying the incident angle of the X-ray w.r.t. to the c-axis to 10 deg, it is possible to excite other nuclear transitions 
which have different energies due to different coupling to the electronic degrees of freedom 
and to allow us to extract the the isomer (electrical monopole) shift $\delta$ and the electrical quadrupole splitting $E_Q$ shifts (see Section \ref{b2} and \ref{b3}).
For the two different charge states of Yb ions in $\beta$-YbAlB$_4$, these values were estimated to be $\delta = 0.27$ mm/s and $E_Q = -2.3$ mm/s 
and $\delta=-0.72$ mm/s and $E_Q=0$ mm/s for two selected $^{174}$Yb nuclear transitions, respectively. 
Since the total angular momentum $J$ is a good quantum number, the electrical quadrupole interaction derives from the nonspherical $4f$ electron distribution \cite{Winkelmann98}.
The second of these transitions with $E_Q=0$ corresponds to the nuclear transitions of the isotropic $(J=0)$ Yb$^{2+}$ ion with $^1S_0$. 
The difference in $\delta$ between these nuclear transitions is comparable with that between the Yb$^{3+}$ and Yb$^{2+}$ ionic state reported 
by the conventional $^{174}$Yb M$\ddot{\text{o}}$ssbauer spectroscopy \cite{Henning71}.
Furthermore, these extracted $\delta$ values are compatible with the electron densities at the Yb nuclei for 
the Yb$^{2+}$ and Yb$^{3+}$ ionic states in $\beta$-YbAlB$_4$ calculated by using the full-potential Korringa-Kohn-Rostoker method \cite{Ogura06,Akai17}.  
Thus, the selected nuclear transition with $\delta=0.27$ mm/s and $E_Q=-2.3$ mm/s 
{is identified as the nuclear transition of the Yb$^{3+}$ configuration with $^2F_{7/2}$.

Recent experimental results and theoretical predictions find a strongly anisotropic hybridization between the Yb $4f$ electrons and 
conduction electrons in $\beta$-YbAlB$_4$, suggsting its pure $J_z=\pm 5/2$ ground doublet states
\cite{Matsumoto11,Nevidomskyy09,Matsumoto15}, attributed to the crystalline-electric-field (CEF) ground state of the Yb $4f$ electrons.  
Interestingly, the extracted $E_Q$ value indicates that the ground Kramers doublet is almost pure $J_z=\pm 5/2$ states (see Section IV-a). 
The CEF ground state of the Yb$^{3+}$ $4f$ electrons infered from our M$\ddot{\text{o}}$ssbauer studies is thus fully consistent with 
both the theoretical material modeling, the single-ion crystal field analysis \cite{Nevidomskyy09,Ramires12} and recent Core-level spectroscopy measurements\cite{Kuga19}. }}

\end{itemize}

\section{Planckian Dissipation in $\beta$-YbAlB$_4$}
\begin{figure}
\centering
\includegraphics[width=0.75\linewidth]{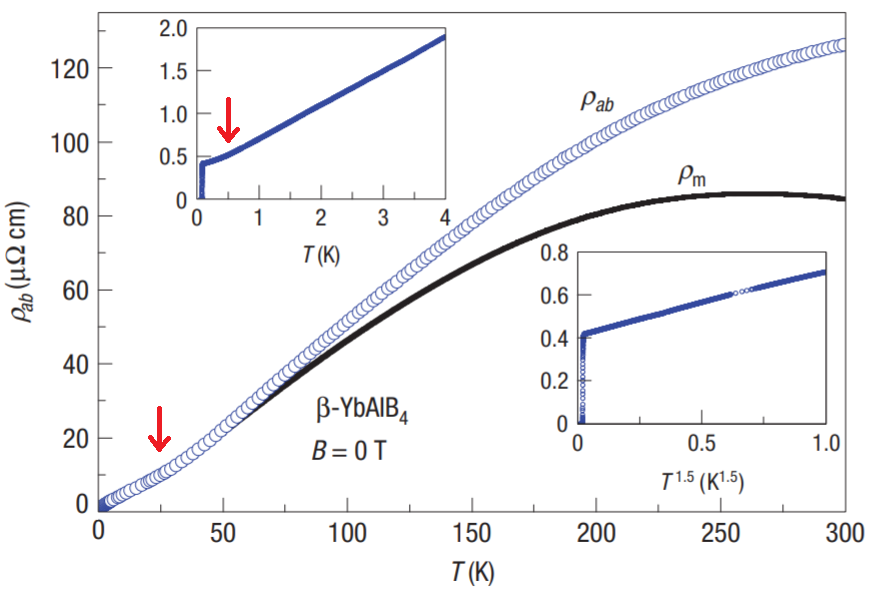}
\caption{\small The resistivity of $\beta$-YbAlB$_4$ showing $\rho\sim T$ for over the range  $0.5$-$25$K (as labeled by arrows) and a slightly different slope over the range $25$K to $150$K, from \cite{Nakatsuji08}.
}\label{fig:res}
\end{figure}

We now establish that the strange metal regime of
$\beta$-YbAlB$_4$ lies in the
Planckian dissipation regime. The resistivity in the temperature range
between 0.25K and 25K reported in \cite{Nakatsuji08}, [Fig.\,\label{fig:res}] has a linear temperature dependence. From these data, we  extract 
\begin{equation}
\rho=\rho_0+A_{\rho}T, \qquad \text{where} \qquad
A_{\rho}=0.4\mu\Omega cm/K \qquad 0.5K < T < 25K.
\end{equation}
We use the Drude formula
\begin{equation}
 \sigma=\frac{ne^2\tau_{tr}}{m} \qquad \to \qquad
\frac{1}{\tau_{tr}}=\frac{ne^2\rho}{m} 
\end{equation} 
to find the scattering rate.  
The effective mass can be eliminated using Fermi temperature:
\begin{equation}
\Big(\frac{1}{\tau_{tr}}\Big)_{\rm 3D}=\frac{k_BT}{\hbar}\times
\zeta, \qquad \zeta=\frac{A_\rho^{\square}T_F}{h/2e^2}, 
\end{equation}
In $\beta$-YbAlB$_4$, the lattice parameters are $a=0.73nm$, $b=0.93nm$ and
$c=0.35nm$. Therefore, the system is relatively three-dimensional. Nevertheless, we have cast the scattering rate in terms of an effective ``sheet density'' $A_\rho^{\square}\equiv A_\rho (3\pi^2n)^{1/3}$ so that the results can be compared to those of Cuprate \cite{Legros2019}. We assume that 
each unit cell contributes roughly one electron to the charge
density, i.e. $n\sim 6\times 10^{27} m^{-3}$. This is consistent with our measurements the Hall effect \cite{OFarrell2012}. The Fermi temperature is
obtained directly from the heat capacity $C=\gamma T$ with
$\gamma={\pi^2}k_B^2\nu(\epsilon_F)/3$ which can be written as 
\begin{equation}
\gamma_{\rm 3D}=\frac{\pi^2}{2}k_B\frac{nV}{T_F}
\end{equation} 
Considering $\gamma=0.1$ J/K$^2$/mole \cite{Nakatsuji08}, we have to set
$k_BnV=k_BN_A=8.3$, finding $T_F=273$K in the 3D model, small but plausible values for a heavy-fermion
metal. Finally, 
\begin{equation}
\Big(\frac{1}{\tau_{\rm tr}}\Big)_{\rm 3D}=\frac{k_BT}{\hbar}\times
\zeta, \qquad \zeta=0.4
\end{equation}
While the precise value of $\zeta$ depends on the assumptions of the model, the important point is that it is a constant of $\zeta\sim O(1)$. Above 25K the increased linear resistivity coefficient corresponds to $\tau^{-1}_{tr}= 0.63\tau^{-1}_{Pl}$ (25K$<T<$150K) in terms of the Planckian scattering rate $\tau^{-1}_{Pl}=k_BT/\hbar$. As discussed in the paper, the change in the slope of linear resistivity is correlated with the change in the dynamics of phonons. The appearance of a transport scattering rate of order $k_{B}T/\hbar$
is the hallmark of a Planckian metal.

\section{Supporting Theoretical Discussion}

\subsection{Kondo Breakdown}
The Kondo breakdown scenario can be captured by a number of approaches, including the slave-boson hybridization  \cite{Pepin07}, 
the numerical renormalization group studies of the pseudogap single-impurity Anderson model \cite{Pixley2012} 
or Schwinger boson large-N study of the Kondo lattice \cite{Komijani19}. 
Fig.\,\ref{Fig:Theory}(A) shows  the divergence of the charge susceptibility at a Kondo breakdown transition studied with the latter method. 
Within a simple Anderson model

\begin{equation}
H_{\rm Anderson}=H_{c}+U_0(n_d)^2+\sum_{\sigma=\uparrow,\downarrow}V_0(c^\dagger_{k\sigma} d_\sigma+h.c.)+\mu_d n_d, \qquad n_d=\sum_{\sigma=\uparrow\downarrow}d^\dagger_\sigma d_\sigma,
\end{equation}
the hybridization and the interaction are renormalized $V_0\to V=ZV_0$ and $U_0\to 0$ due to the Kondo effect, where the wavefunction renormalization $Z$ is given by
\begin{equation}
Z=[1-\partial_\omega\Sigma_f(\omega)]^{-1}.
\end{equation}
At the Kondo-breakdown transition, the localization of the $d$ electrons means that their effective mass diverges and the $Z\to 0$, 
further renormalizing the hybrdization to zero $V\to 0$ \cite{Pepin07,Pixley2012,Komijani19}. 
The fluctuation time-scale is associated with the hybrdization $V=ZV_0$
\begin{equation}
\tau^{-1}_f\sim \pi\rho V^2
\end{equation}
where $\rho$ is the density of states of the conduction electrons at the Fermi energy.
Therefore, the renormalization $V\to 0$ at the Kondo break-down leads to slow charge fluctuations. 

Fig.\,\ref{Fig:Theory}(A) shows the static charge susceptibility vs. temperature across a Kondo breakdown transition 
between a FL phase and a spin-liquid, computed using Schwinger boson approach \cite{Komijani19}. 
The $\chi_\rho\sim 1/T$ at the transition indicates to the emergence of a sharp zero-frequency peak in the dynamical charge susceptibility 
$-\beta \chi''_\rho(\omega)/\omega$, as shown in the inset of Fig.\,\ref{Fig:Theory}(B). 
Fig.\,\ref{Fig:Theory}(B) also shows the result of this critical charge susceptibility on the M$\ddot{\text{o}}$ssbauer line shape 
where a single peak at high temperature is splitted into two peaks proportional to $\alpha^2$.

In spite of qualitative agreement, the splitting observed in the paper is too large compared to the temperature and cannot be entirely accounted for 
by the mechanism discussed here. 

The resolution of this paradox involves the role of electron-phonon coupling: The phonons respond to the charge accumulation by dressing the charges into polarons, 
thereby additionally renormalizing the hybridization and thus the width of the dynamical charge susceptibility. 

\subsection{Phonons and Polarons}
The physical essence can be captured by a very simplified model
\begin{equation}
H=H_{\rm Anderson}+\sum_q[\omega_q a_q^\dagger a_q+g_q (a_q^\dagger+a_q)n_d].
\end{equation}
The normal phononic modes are related to the crystal displacement and momentum as
\begin{equation}
\Delta z=\sum_q\sqrt{\frac{\hbar}{2m\omega_q}}(a_q+a^\dagger_q), \qquad p_z=\sum_q i\sqrt{\frac{\hbar m\omega_q}{2}}(a^\dagger_q-a_q)\label{eq15}
\end{equation}
The phonon mode is approximated by the Debye model with a characteristic energy scale $k_{\rm B} \Theta_{\rm D}$ 
and $g_q$ is the coupling of the phonons $\omega_q$ to the occupation of the $f$-electrons. 
The effect of polarons is non-perturbative in $g_q$ and can be best captured by the so-called Lang-Firsov unitary transformation \cite{langfirsov}
\begin{equation}
U=\exp[\sum_qu_q  (a_q^\dagger -a_q)n_d],
\end{equation}
where $n_d=\sum_\sigma d^\dagger_\sigma d_\sigma$. As a result of this transformation, we have
\begin{equation}
V\to Ve^{\sum_qu_q(a_q^\dagger-a_q)}, \qquad g_q\to g_q-u_q\omega_q, \qquad U\to U-\sum_q\omega_q u_q^2.
\end{equation}
Therefore, choosing $u=g_q/\omega_q$ eliminates the electron-phonon coupling. But the hybridization is renormalized:
\begin{equation}
V\to V\langle e^{\sum_qu_q(a_q^\dagger-a_q)}\rangle=Ve^{-\sum_qu_q^2(n_q+1/2)}.\label{eq17}
\end{equation}
Here, $n_q$ is the number of bosons $\omega_q$ and it is given by the Bose-Einstein distribution in a thermal state $n_q=n_B(\omega_q)$ 
where $n_B(\omega)=[1+e^{\omega/T}]^{-1}$ at temperature $T$. 
At low-temperature thermal fluctuations are suppressed $n_B(\omega_p)=0$, and only the quantum fluctuations contribute 
to the suppression of the renormalized matrix element of the hybridization $V$. 
Note that as a result of the unitary transformation
\begin{equation}
a_q^\dagger a_q\to U^\dagger a_q^\dagger a_q U=(a_q^\dagger-u_qn_d)(a_q-u_qn_d)=a_q^\dagger a_q-u_qn_d(a_q^\dagger+a_q)+u^2_qn_d^2.\label{eq19}
\end{equation}
Therefore, the phonon occupancy departs from the thermal equilibrium to $n_q=\langle a^\dagger_q a_q\rangle=n_B(\omega_q)+u_q^2n_d^2$. 
For simplicity, we limit the discussion to the transition between empty ($n_d=0$) and singly occupied ($n_d=1$) infinite-U Anderson model. 
Therefore, the quantum fluctuation contribution $N_p\equiv \sum_qu^2_q$ in the exponent of Eq.\,(\ref{eq17}) can be interpreted 
as the total number of phonons excited by the change in the charge $n_d=0\to n_d=1$ and involved in the polaron. 
This leads to the additional slowing-down of the charge fluctuations
\begin{equation}
\tau^{-1}_f\sim \pi\rho V^2e^{-2N_p}.
\end{equation}
This discussion and the interpretation of $N_p$ as the number of phonons excited by the charge goes back to \cite{Sherington76,Hewson1979}.



\clearpage
\begin{figure}
 \begin{center}
  \includegraphics [width=0.7\linewidth]  {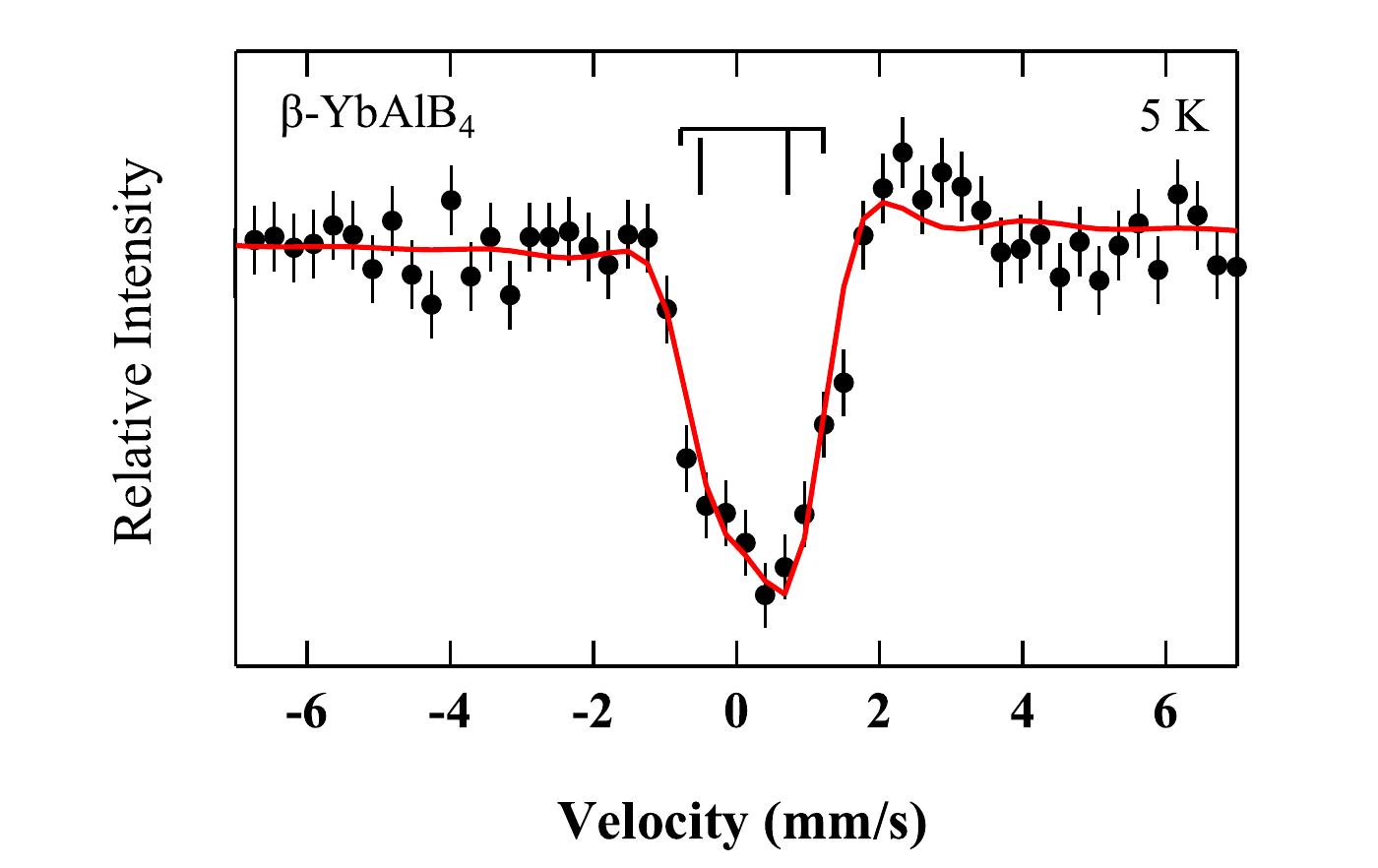}
\vspace*{+1.3cm}
   \caption{
Synchrotron-radiation-based $^{174}$Yb M$\ddot{\text{o}}$ssbauer spectrum of $\beta$-YbAlB$_4$ at 5K.
The $c$-axis of the single crystalline $\beta$-YbAlB$_4$ samples was tilted though 10 deg from the incident X-ray. 
The closed circles with error bar indicate the observed spectrum and the red solid line presents the analytical spectrum. 
The extracted subspectra for two Yb states are shown by the bar-diagrams. 
The relative intensities of the bar-diagrams represent the evaluated excitation probabilities for the allowed $^{174}$Yb  nuclear transitions 
in the experimental conditions.
}
\label{Fig:Fig_Moss10}
\end{center}
\end{figure}

\clearpage

\begin{figure}
 \begin{center}
  \includegraphics [width=0.7\linewidth]  {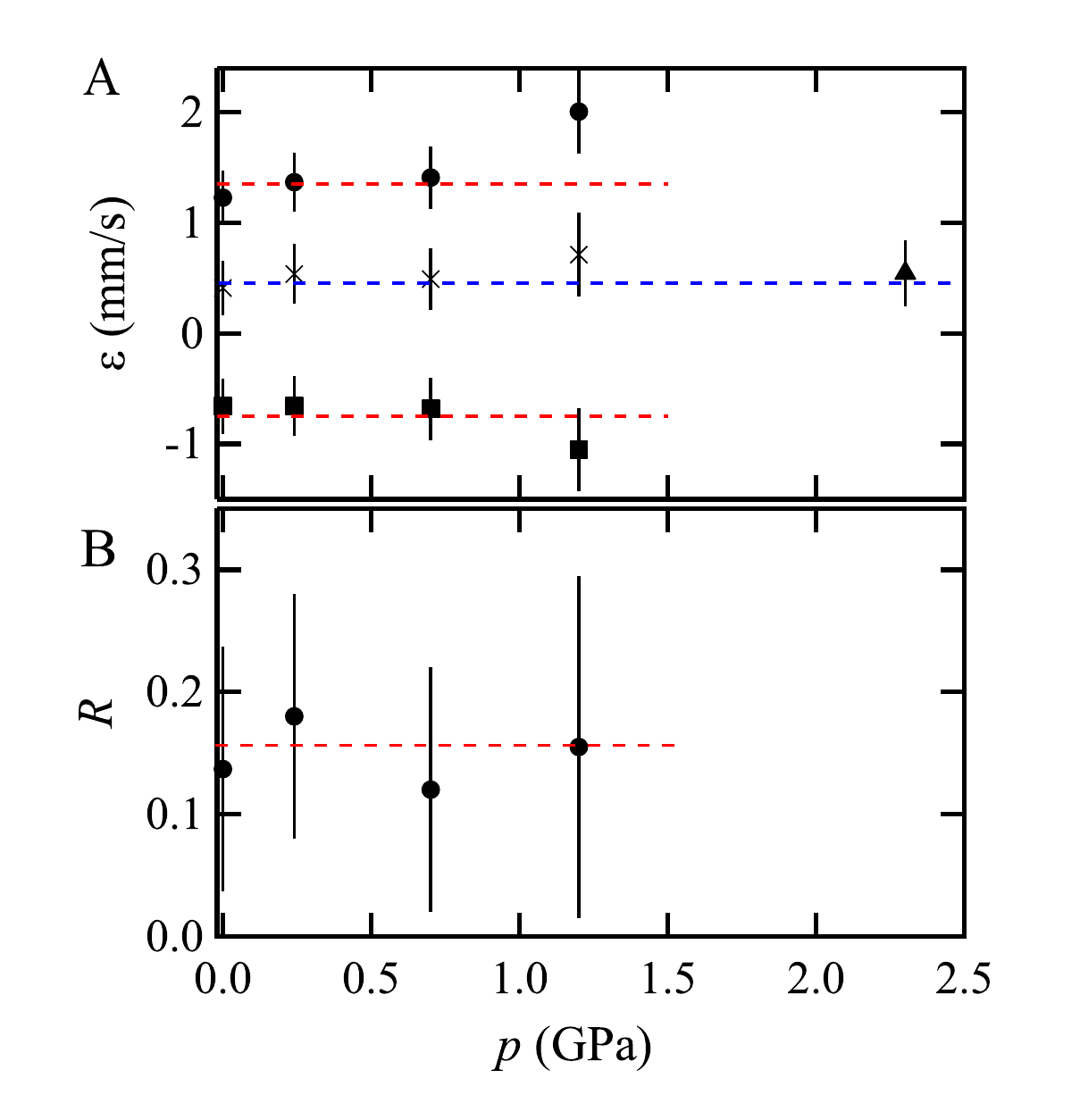}
\vspace*{+1.3cm}
   \caption{
(A) Refined and evaluated energies $\varepsilon$ for the nuclear transitions and (B) difference in intensity $R$ between these two nuclear transitions 
in $\beta$-YbAlB$_4$ at 2K as functions of pressure. 
In (A), the solid circles and squares with error bar presents the refined $\varepsilon$ for two selected $I_g = 0 \rightarrow I_e^z = \pm 1$ nuclear transitions, 
respectively, below 1.2GPa.
The crosses represent the weighted average $\varepsilon$ values of two nuclear transitions below 1.2GPa. 
The solid triangle with error bar indicates the refined $\varepsilon$ for the single absorption component 
in the SR-based $^{174}$Yb M$\ddot{\text{o}}$ssbauer spectrum of $\beta$-YbAlB$_4$ at 2.3GPa and 2K.
The red and blue broken lines are visual guides.
}
\label{Fig:Fig_Evsp}
\end{center}
\end{figure}

\begin{figure}[h!]
\centering
  \includegraphics [width=0.8\linewidth]  {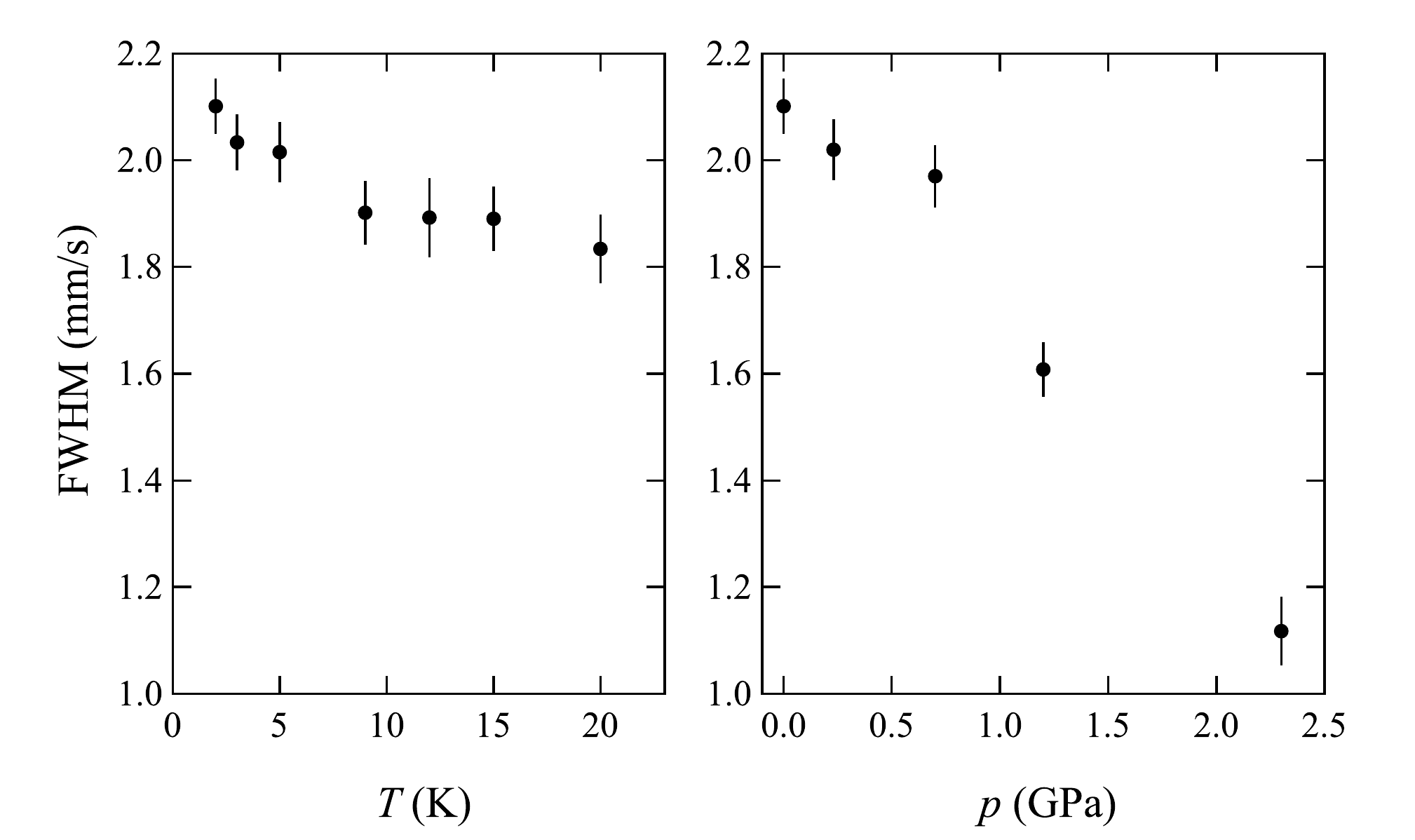}
\vspace*{+0.5cm}
   \caption{
 Evaluated FWHM of the dip in the observed spectra.
}\label{Fig:Fig_3c_in}
\end{figure}

\begin{figure}[h!]
\includegraphics[width=1\linewidth]{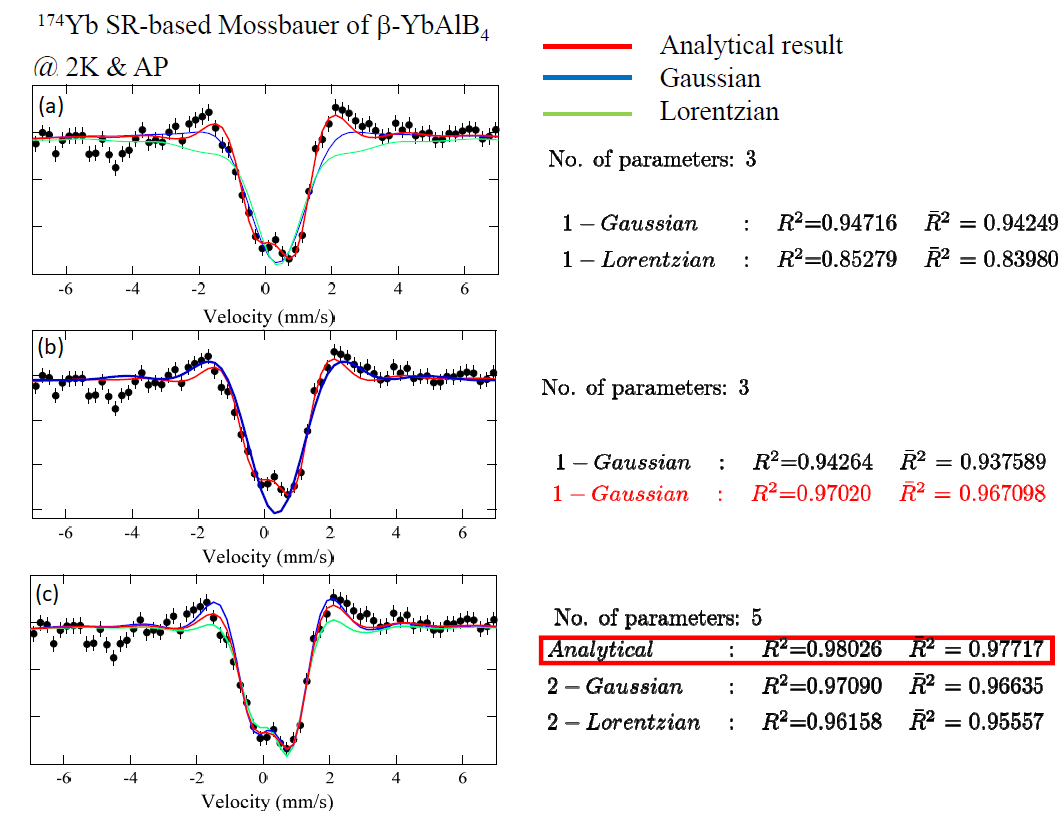}
\caption{Various fits to the experimental lineshape at 2K and ambient pressure.}\label{fig4}
\end{figure}

\begin{figure}
 \begin{center}
  \includegraphics [width=\linewidth] {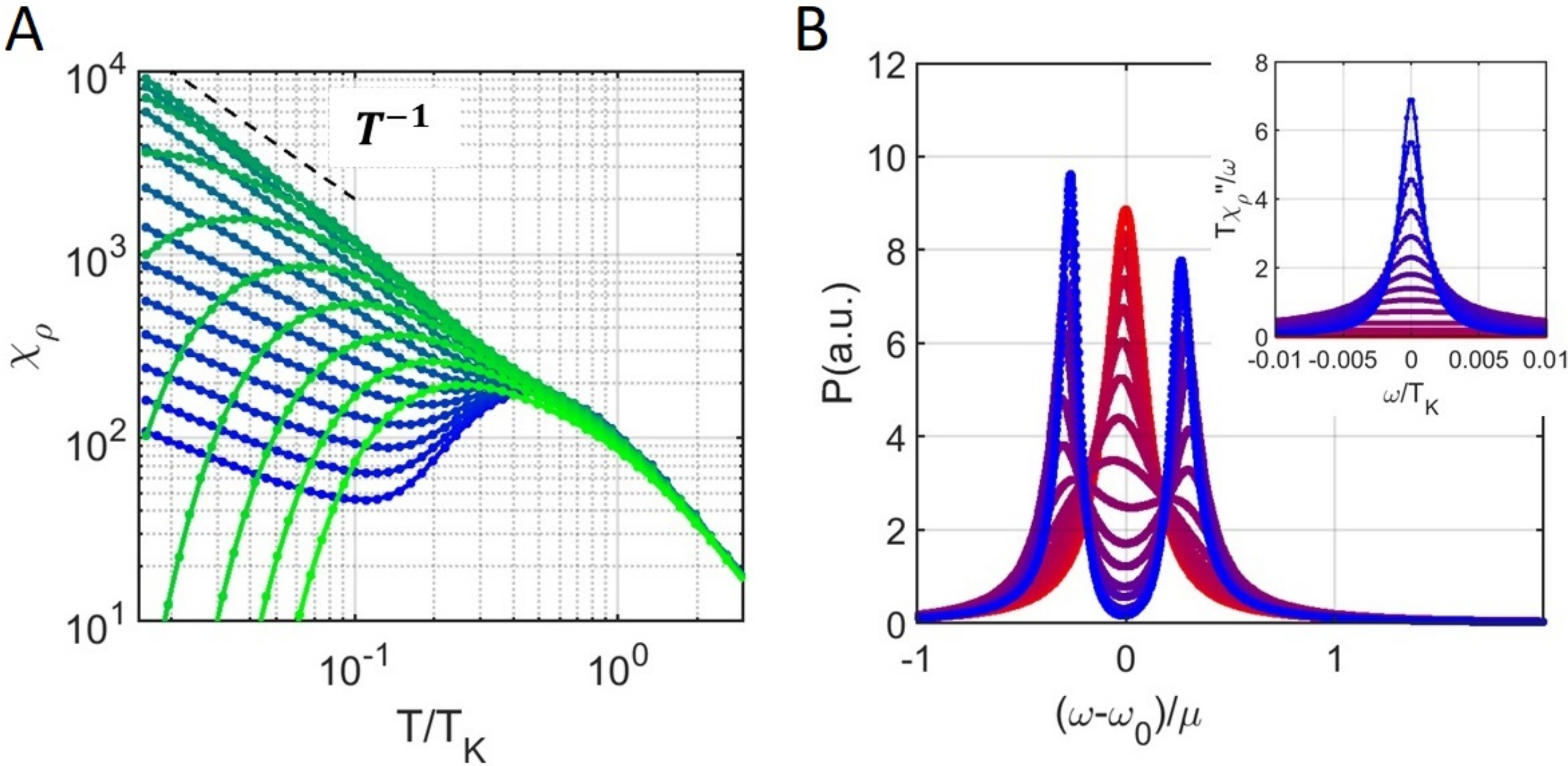}
\vspace*{+1.3cm}
   \caption{
(A) The static charge susceptibility as a function of $T$ for various values of $T_K/J_H$ shows the divergence of
the charge susceptibility at the Kondo breakdown QCP between a FL (blue) and a spin-liquid (green). (B) Inset: The dynamical charge
susceptibility vs. frequency at the QCP, shows a sharp feature developing
as the temperature is lowered from red to blue. Main panel: Consequent splitting of the M\"ossbauer lineshape at low-temperature as a result of the critical charge fluctuation.}
\label{Fig:Theory}
\end{center}
\end{figure}

\clearpage
\bibliography{YbAlB4Ref}
\bibliographystyle{science}